\documentclass[prd,twocolumn,floatfix,nofootinbib]{revtex4}
\usepackage{amssymb}
\usepackage{amsmath}
\usepackage{graphicx,subfigure,color,dcolumn,booktabs,bm}
\usepackage{longtable,lscape}
\usepackage{txfonts}
\usepackage{overpic}
\usepackage{indentfirst}
\usepackage{cases}
\usepackage{multirow}
\usepackage{epstopdf}
\usepackage{ulem}
\usepackage[colorlinks,
            citecolor=blue,
            anchorcolor=red,
            menucolor=red,
            linkcolor=red,
            filecolor=red,
            runcolor=red,
            urlcolor=blue,
            frenchlinks=false]{hyperref}
\graphicspath{{Figures/}} %
\allowdisplaybreaks

\begin{document}

\title{Mass spectra of hidden heavy-flavor tetraquarks with two and four heavy
quarks}

\author{Ting-Qi Yan$^{1}$}
\email{Yantingqi@outlook.com}
\author{Wen-Xuan Zhang$^{1}$}
\email{zhangwx89@outlook.com}
\author{Duojie Jia$^{1,2}$
\footnote{Corresponding author}}
\email{jiadj@nwnu.edu.cn}

\affiliation{ $^1$Institute of Theoretical Physics, College of Physics and
Electronic Engineering, Northwest Normal University, Lanzhou 730070, China \\
$^2$Lanzhou Center for Theoretical Physics, Lanzhou University, Lanzhou, 730000, China \\ }


\begin{abstract}Inspired by the observation of the $X(6900)$ by LHCb and the $X(6600)$ (with mass $6552\pm 10$ $\pm 12$ MeV) recently by CMS and ATLAS experiments of the LHC in the di-$J/\Psi $ invariant mass spectrum, we systemically study masses of all ground-state configurations of the hidden heavy-flavor tetraquarks $q_{1}Q_{2}\bar{q}_{3}\bar{Q}_{4}$ and $Q_{1}Q_{2}\bar{Q}_{3}\bar{Q}_{4}$ ($Q=c,b$;$q=u,d,s$) contaning two and four heavy
quarks in the MIT bag model with chromomagnetic interaction and
enhanced binding energy. Considering color-spin mixing due to chromomagnetic interaction, our mass computation indicates that the observed $X(6600)$ is likely to be the $0^{++}$ ground states of hidden-charm tetraquark $cc\bar{c}\bar{c}$ with computed masses $6572$ MeV, which has a $0^{++}$ color partner around
$6469$ MeV. The fully bottom system of tetraquark $bb\bar{b}\bar{b}$ has masses of 19685 MeV and 19717 MeV for the the $0^{++}$ ground states. Further computation is given to the tetraquark systems $sc\bar{s}\bar{c}$, $sb\bar{s}\bar{b}$, $cb\bar{c}\bar{b}$, $nc\bar{n}\bar{c}$ and $nb\bar{n}\bar{b}$, suggesting that the $Z_{c}(4200)$ is the tetraquark $nc\bar{n}\bar{c}$ with $J^{PC}=1^{+-}$. All of these tetraquarks are above their lowest thresholds of two  mesons and unstable against the strong decays.

{\normalsize PACS number(s):12.39Jh, 12.40.Yx, 12.40.Nn}

\end{abstract}
\maketitle
\date{\today}

\section{Introduction}

\label{intro} All known strongly interacting particles (mesons and baryons)
could be classified as bound states made of a quark-antiquark pair or three
quarks for a long time based on the conventional scheme of the quark model
by Gell-Mann\cite{Gell-Mann:1964ewy} and Zweig\cite{Zweig:1964ruk}.
Meanwhile, they also suggested possible existence of the hadron states of
multiquarks like tetraquarks (with quark configuration $q^{2}\bar{q}^{2}$)
and pentaquarks ($q^{4}\bar{q}$). In the 1970s, multiquark states (the
exotic light mesons like the $a_{0}$ and $f_{0}$) are calculated by Jaffe
based on the dynamical framework of the MIT bag model \cite%
{Jaffe:1976ig,Jaffe:1976ih}. Despite that multiquarks are considered to be
exotic in the sense that they go beyond the conventional scheme of quark
model, they are, in principle, allowed by the quantum chromodynamics (QCD),
the theory of the strong force that binds quarks into hadrons.

Since observation of the first exotic hadron $X(3872)$ \cite{Belle:2003nnu}
in 2003 by the Belle, many (more than 20) tetraquark candidates have been
observed among charmonium-like or bottomonium-like $XYZ$ states, which
include the charmonium-like states the $Z_{c}(3900)$ \cite{BESIII:2013ris},
the $Z_{c}(4200)$ \cite{Belle:2014nuw}, the $Z_{c}(4430)$ \cite%
{Belle:2007hrb,Belle:2013shl,LHCb:2014zfx,LHCb:2015sqg}. Some of the
observed $XYZ$ states, like the charged state $Z_{c}(3900)$ \cite%
{BESIII:2013ris}, are undoubtedly exotic. In 2020, a candidate of fully
charm tetraquark, the $X(6900)$, has been observed by LHCb in the di-$J/\Psi
$ invariant mass spectrum around the mass of $6905$ MeV, which is later
confirmed by CMS and ATLAS of the LHC at CERN \cite%
{Bouhova-Thacker:2022vnt,Zhang:2022toq,LHCb:2020bwg}. Meanwhile in the same
di-$J/\Psi $ invariant mass spectrum, a new structure, the $X(6600)$, are
also found by CMS with mass of $6552\pm 10$ $\pm 12$ MeV, which is very
likely to be the fully charm tetraquark.

The purpose of this work is to use the MIT bag model with enhanced binding
energy to systemically study the ground-state masses of the hidden
heavy-flavor tetraquarks containing two or four heavy quarks. Based on
color-spin wavefunctions constructed for the hidden heavy-flavor
tetraquarks, we solve the bag model and diagonalize the chromomagnetic
interaction (CMI) to take into account the possible color-spin mixing of the
states with same quantum numbers. We find that the computed masses of the
fully charmed tetraquark $cc\bar{c}\bar{c}$ is in a good agreement with the
mass measurement by the CMS experiment\cite{Zhang:2022toq}. Further mass
computation is performed for hidden heavy-flavor systems of the tetraquarks $bb%
\bar{b}\bar{b}$, $cb\bar{c}\bar{b}$, $sc\bar{s}\bar{c}$, $sb\bar{s}\bar{b}$,
$nc\bar{n}\bar{c}$, $nb\bar{n}\bar{b}$, with a suggestion that the particle $%
Z_{c}(4200)$ reported by \cite{Belle:2014nuw} is likely to be the
hidden-charm tetraquark $nc\bar{n}\bar{c}$ with $J^{PC}=1^{+-}$.

In Section 2, we present the allowed wavefunctions of hidden heavy-flavor
tetraquarks with two or four heavy quarks. In Section 3, We describe the
framework of MIT bag model to be used in this work. The mass matrix
evaluation for the CMI and its diagonalization are detailed in section 4.
The masses of the hidden heavy-flavor tetraquarks are computed numerically for the
systems ($cc\bar{c}\bar{c}$, $bb\bar{b}\bar{b}$), $cb\bar{c}\bar{b}$, ($sc%
\bar{s}\bar{c}$, $sb\bar{s}\bar{b}$) and ($nc\bar{n}\bar{c}$, $nb\bar{n}\bar{%
b}$) in Section 5. We end with conclusions and remarks in Section 6.

\section{Wavefunctions of hidden-flavor tetraquarks}

\label{sec:1}

We consider hidden heavy-flavor tetraquarks containing two or four heavy
quarks($q_{1}Q_{2}\bar{q}_{3}\bar{Q}_{4}$ and $Q_{1}Q_{2}\bar{Q}_{3}\bar{Q}%
_{4}$, $Q=c,b$, $q=u,d,s$), which include seven flavor combinations of four
quark systems: $cc\bar{c}\bar{c}$, $bb\bar{b}\bar{b}$, $sc\bar{s}\bar{c}$, $%
sb\bar{s}\bar{b}$, $cb\bar{c}\bar{b}$, $nc\bar{n}\bar{c}$, $nb\bar{n}\bar{b}$%
, with $n=u$, $d$. In this section, we describe the wavefunctions of the
hidden-flavor tetraquarks in the flavor and the color-spin space.

In the flavor space, we utilized $\delta _{12}^{S}$ if $q_{1}q_{2}$ is
symmetric and $\delta _{12}^{A}\equiv 1-\delta _{12}^{A}$ if $q_{1}q_{2}$ is
antisymmetric to restrict the wavefunction. If the wavefunction has no
flavor symmetry (beyond the isospin symmetry $SU(2)_{I}$) under the exchange
of $q_{1}$ and $q_{2}$, then $\delta _{12}^{S}=\delta _{12}^{A}=1$.

In color space, the hidden heavy-flavor tetraquark $q_{1}q_{2}\bar{q}_{3}%
\bar{q}_{4}$ can be in two color states: $6_{c}\otimes \bar{6}_{c}$ and $%
\bar{3}_{c}\otimes 3_{c}$, with the respective wave functions (superscript
stands for color representation),
\begin{equation}
\phi _{1}^{T}=\left\vert {\left( q_{1}q_{2}\right) }^{6}{\left( \bar{q}_{3}%
\bar{q}_{4}\right) }^{\bar{6}}\right\rangle ,\quad \phi _{2}^{T}=\left\vert {%
\left( q_{1}q_{2}\right) }^{\bar{3}}{\left( \bar{q}_{3}\bar{q}_{4}\right) }%
^{3}\right\rangle .  \label{colorT}
\end{equation}%
With the help of the color $SU(3)_{c}$ symmetry, one can write the two
configurations $\phi _{1,2}^{T}$ here in terms of the fundamental
representations, i.e., of the color bases $c_{n}=|r\rangle $, $|b\rangle $
and $|g\rangle $ of the $SU(3)_{c}$ group (see Appendix A).

In the spin space, there are six states of a tetraquark state allowed
(Appendix A), with the wavefunctions (subscript stands for spin),
\begin{equation}
\begin{aligned} \chi_{1}^{T}={\left| {\left(q_{1}q_{2}\right)}_{1}
{\left(\bar{q}_{3}\bar{q}_{4}\right)}_{1} \right\rangle}_{2}, \quad
\chi_{2}^{T}={\left| {\left(q_{1}q_{2}\right)}_{1}
{\left(\bar{q}_{3}\bar{q}_{4}\right)}_{1} \right\rangle}_{1}, \\
\chi_{3}^{T}={\left| {\left(q_{1}q_{2}\right)}_{1}
{\left(\bar{q}_{3}\bar{q}_{4}\right)}_{1} \right\rangle}_{0}, \quad
\chi_{4}^{T}={\left| {\left(q_{1}q_{2}\right)}_{1}
{\left(\bar{q}_{3}\bar{q}_{4}\right)}_{0} \right\rangle}_{1}, \\
\chi_{5}^{T}={\left| {\left(q_{1}q_{2}\right)}_{0}
{\left(\bar{q}_{3}\bar{q}_{4}\right)}_{1} \right\rangle}_{1}, \quad
\chi_{6}^{T}={\left| {\left(q_{1}q_{2}\right)}_{0}
{\left(\bar{q}_{3}\bar{q}_{4}\right)}_{0} \right\rangle}_{0}. \end{aligned}
\label{spinT}
\end{equation}

Based on the Pauli's principle, one can construct twelve color-spin
wavefunctions for the lowest S-wave (in coordinate space) tetraquarks:
\begin{equation}
\begin{aligned} \phi_{1}^{T}\chi_{1}^{T}={\left|
{\left(q_{1}q_{2}\right)}_{1}^{6}
{\left(\bar{q}_{3}\bar{q}_{4}\right)}_{1}^{\bar{6}} \right\rangle}_{2}
\delta_{12}^{A}\delta_{34}^{A}, \\ \phi_{2}^{T}\chi_{1}^{T}={\left|
{\left(q_{1}q_{2}\right)}_{1}^{\bar{3}}
{\left(\bar{q}_{3}\bar{q}_{4}\right)}_{1}^{3} \right\rangle}_{2}
\delta_{12}^{S}\delta_{34}^{S}, \\ \phi_{1}^{T}\chi_{2}^{T}={\left|
{\left(q_{1}q_{2}\right)}_{1}^{6}
{\left(\bar{q}_{3}\bar{q}_{4}\right)}_{1}^{\bar{6}} \right\rangle}_{1}
\delta_{12}^{A}\delta_{34}^{A}, \\ \phi_{2}^{T}\chi_{2}^{T}={\left|
{\left(q_{1}q_{2}\right)}_{1}^{\bar{3}}
{\left(\bar{q}_{3}\bar{q}_{4}\right)}_{1}^{3} \right\rangle}_{1}
\delta_{12}^{S}\delta_{34}^{S}, \\ \phi_{1}^{T}\chi_{3}^{T}={\left|
{\left(q_{1}q_{2}\right)}_{1}^{6}
{\left(\bar{q}_{3}\bar{q}_{4}\right)}_{1}^{\bar{6}} \right\rangle}_{0}
\delta_{12}^{A}\delta_{34}^{A}, \\ \phi_{2}^{T}\chi_{3}^{T}={\left|
{\left(q_{1}q_{2}\right)}_{1}^{\bar{3}}
{\left(\bar{q}_{3}\bar{q}_{4}\right)}_{1}^{3} \right\rangle}_{0}
\delta_{12}^{S}\delta_{34}^{S}, \\ \phi_{1}^{T}\chi_{4}^{T}={\left|
{\left(q_{1}q_{2}\right)}_{1}^{6}
{\left(\bar{q}_{3}\bar{q}_{4}\right)}_{0}^{\bar{6}} \right\rangle}_{1}
\delta_{12}^{A}\delta_{34}^{S}, \\ \phi_{2}^{T}\chi_{4}^{T}={\left|
{\left(q_{1}q_{2}\right)}_{1}^{\bar{3}}
{\left(\bar{q}_{3}\bar{q}_{4}\right)}_{0}^{3} \right\rangle}_{1}
\delta_{12}^{S}\delta_{34}^{A}, \\ \phi_{1}^{T}\chi_{5}^{T}={\left|
{\left(q_{1}q_{2}\right)}_{0}^{6}
{\left(\bar{q}_{3}\bar{q}_{4}\right)}_{1}^{\bar{6}} \right\rangle}_{1}
\delta_{12}^{S}\delta_{34}^{A}, \\ \phi_{2}^{T}\chi_{5}^{T}={\left|
{\left(q_{1}q_{2}\right)}_{0}^{\bar{3}}
{\left(\bar{q}_{3}\bar{q}_{4}\right)}_{1}^{3} \right\rangle}_{1}
\delta_{12}^{A}\delta_{34}^{S}, \\ \phi_{1}^{T}\chi_{6}^{T}={\left|
{\left(q_{1}q_{2}\right)}_{0}^{6}
{\left(\bar{q}_{3}\bar{q}_{4}\right)}_{0}^{\bar{6}} \right\rangle}_{0}
\delta_{12}^{S}\delta_{34}^{S}, \\ \phi_{2}^{T}\chi_{6}^{T}={\left|
{\left(q_{1}q_{2}\right)}_{0}^{\bar{3}}
{\left(\bar{q}_{3}\bar{q}_{4}\right)}_{0}^{3} \right\rangle}_{0}
\delta_{12}^{A}\delta_{34}^{A}. \end{aligned}  \label{colorspinT}
\end{equation}%
We choose these wavefunctions to be the bases (the first approximation) of
the tetraquark eigenstates for which the chromomagnetic interaction (CMI)
are ignored. We are going to employ these bases to take into account the
chromomagnetic mixing due to the CMI. For example, for the $J^{PC}=0^{++}$
state of the $cc\bar{c}\bar{c}$ tetraquark, one can write two bases of the
wavefunctions $\phi _{2}^{T}\chi _{3}^{T}$, $\phi _{1}^{T}\chi _{6}^{T}$ in
Eq. (\ref{colorspinT}) as a zero-order approximation, which satisfy the
required symmetry in the color-spin space and can lead to mixing of the
color-spin states when the CMI added.

\renewcommand{\tabcolsep}{1.2cm} \renewcommand{\arraystretch}{2.2}
\begin{table*}[tbh]
\caption{Allowed state mixing of the hidden heavy-flavor tetraquarks due to
chromomagnetic interaction.}
\label{tab:mixed states}%
\begin{tabular}{lcc}
\hline\hline
\textrm{State} & $J^{PC}$ & Allowed states for mixing \\ \hline
$QQ\bar{Q}\bar{Q}$ & $0^{++}$ & $\left( \phi _{2}^{T}\chi _{3}^{T},\phi
_{1}^{T}\chi _{6}^{T}\right) $ \\
& $1^{+-}$ & $\left( \phi _{2}^{T}\chi _{2}^{T}\right) $ \\
& $2^{++}$ & $\left( \phi _{2}^{T}\chi _{1}^{T}\right) $ \\
$Q{Q}^{\prime }\bar{Q}\bar{Q}^{\prime }$,$qQ\bar{q}\bar{Q}$ & $0^{++}$ & $%
\left( \phi _{2}^{T}\chi _{3}^{T},\phi _{2}^{T}\chi _{6}^{T},\phi
_{1}^{T}\chi _{3}^{T},\phi _{1}^{T}\chi _{6}^{T}\right) $ \\
& $1^{++}$ & ($\frac{1}{\sqrt{2}}\left( \phi _{2}^{T}\chi _{4}^{T}+\phi
_{2}^{T}\chi _{5}^{T}\right) ,\frac{1}{\sqrt{2}}\left( \phi _{1}^{T}\chi
_{4}^{T}+\phi _{1}^{T}\chi _{5}^{T}\right) $) \\
& $1^{+-}$ & ($\phi _{2}^{T}\chi _{2}^{T},\frac{1}{\sqrt{2}}\left( \phi
_{2}^{T}\chi _{4}^{T}-\phi _{2}^{T}\chi _{5}^{T}\right) ,\phi _{1}^{T}\chi
_{2}^{T},\frac{1}{\sqrt{2}}\left( \phi _{1}^{T}\chi _{4}^{T}-\phi
_{1}^{T}\chi _{5}^{T}\right) $) \\
& $2^{++}$ & $\left( \phi _{2}^{T}\chi _{1}^{T},\phi _{1}^{T}\chi
_{1}^{T}\right) $ \\ \hline\hline
\end{tabular}%
\end{table*}

With respect to given flavor compositions of the tetraquarks, one can write
the allowed color-spin states that may mix due to the CMI for each choice of
the quantum number $J^{PC}$ in Table 1, where $Q^{\prime }$ denoting heavy
quark differing with $Q$. Note that for the flavor composition $QQ\bar{Q}%
\bar{Q}$ with quantum numbers $J^{PC}=1^{+-}$ and $2^{++}$, there is only
one color-spin state for each of them, that is, the $\phi _{2}^{T}\chi
_{2}^{T}$ associated with $1^{+-}$ and $\phi _{2}^{T}\chi _{1}^{T}$
associated with $2^{++}$, for which not mixing occurs in reality.

\section{The MIT bag model}

We use the MIT bag model which includes enhanced binding energy and the CMI
in the interaction correction $\Delta M$. The mass formula for the MIT bag
model is\cite{Zhang:2021yul}
\begin{equation}
M\left( R\right) =\sum_{i}\omega _{i}+\frac{4}{3}\pi R^{3}B-\frac{Z_{0}}{R}%
+\Delta M,  \label{MBm}
\end{equation}%
\begin{equation}
\omega _{i}=\left( m_{i}^{2}+\frac{x_{i}^{2}}{R^{2}}\right) ^{1/2},
\label{freq}
\end{equation}%
with the first term describes (relativistic) kinetic motion of each quark $i$
in tetraquark, the second is the volume energy of bag with bag constant $B$,
the third is the zero-point-energy with coefficient $Z_{0}$ and $R$ the bag
radius to be determined variationally. In Eq. (\ref{freq}), the
dimensionless parameters $x_{i}=x_{i}(mR)$ are related to $R$ through an
transcendental equation
\begin{equation}
\tan x_{i}=\frac{x_{i}}{1-m_{i}R-\left( m_{i}^{2}R^{2}+x_{i}^{2}\right)
^{1/2}}.  \label{transc}
\end{equation}

In Eq. (\ref{MBm}), we denote the sum of the first three terms to be $M(T) $%
. The interaction correction $\Delta M$ includes the enhanced binding energy
$M_{B}$ among the quarks in tetraquark and the mass splitting $M_{CMI} $
corresponding to the CMI:
\begin{equation}
\Delta M=M_{B}+M_{CMI}=\sum_{i<j}B_{ij}+\langle H_{CMI}\rangle ,  \label{dM}
\end{equation}%
where $B_{ij}$ stands for the binding energy\cite%
{Karliner:2014gca,Karliner:2017elp} between quarks $i$ and $j$, described
below at the end of this section, and the chromomagnetic interaction $%
H_{CMI} $ is given by
\begin{equation}
H_{CMI}=-\sum_{i<j}\left( \mathbf{\lambda }_{\mathbf{i}}\cdot \mathbf{%
\lambda }_{j}\right) \left( \mathbf{\sigma }_{\mathbf{i}}\cdot \mathbf{%
\sigma }_{j}\right) C_{ij}.  \label{CMI}
\end{equation}%
where $\mathbf{\lambda }\boldsymbol{_{i}}$ and $\mathbf{\sigma }\boldsymbol{%
_{i}}$ are the Gell-Mann and Pauli matrices of the quark $i$, respectively,
and $C_{ij}$ the CMI coupling parameters, given by\cite{DeGrand:1975cf}

\begin{equation}
C_{ij}=3\frac{\alpha _{s}\left( R\right) }{R^{3}}\bar{\mu}_{i}\bar{\mu}%
_{j}I_{ij},  \label{Cij}
\end{equation}%
with $\alpha _{s}(R)$ is the running coupling given in Ref. \cite%
{Zhang:2021yul}, $\bar{\mu}_{i}$ the reduced magnetic moment of quark $i$,
\begin{equation}
\alpha _{s}(R)=\frac{0.296}{ln\left[ 1+{\left( 0.281R\right) }^{-1}\right] }.
\label{alphaS-mine}
\end{equation}%
\begin{equation}
\bar{\mu}_{i}=\frac{R}{6}\frac{4{\omega _{i}R}+2\lambda _{i}-3}{2{\omega
_{i}R}\left( {\omega _{i}R}-1\right) +\lambda _{i}},  \label{muBari}
\end{equation}%
and
\begin{equation}
I_{ij}=1+2\int_{0}^{R}\frac{dr}{r^{4}}\bar{\mu}_{i}\bar{\mu}_{j}=1+F\left(
x_{i},x_{j}\right) .  \label{Iij}
\end{equation}%
where $\lambda _{i}\equiv {m_{i}R}$. The function $F\left(
x_{i},x_{j}\right) $ is given by
\begin{equation}
\begin{aligned} F\left (x_{i},x_{j} \right )=\left (x_{i} \sin^{2}
x_{i}-\frac{3}{2} y_{i} \right ) ^{-1}\left (
x_{j}\sin^{2}x_{j}-\frac{3}{2}y_{j} \right ) ^{-1} \\ \left \{
-\frac{3}{2}y_{i}y_{j}-2x_{i}x_{j}sin^{2}x_{i}sin^{2}x_{j}+%
\frac{1}{2}x_{i}x_{j}\left [ 2x_{i}Si \left ( 2x_{i}\right ) \right.\right.
\\ \left.\left. +2x_{j}Si\left ( 2x_{j} \right )-\left ( x_{i}+x_{j} \right
)Si\left ( 2\left ( x_{i}+x_{j} \right ) \right ) \right.\right. \\
\left.\left.-\left ( x_{i}-x_{j} \right )Si\left ( 2\left ( x_{i}-x_{j}
\right ) \right ) \right ] \right \} \end{aligned}  \label{Fxixj}
\end{equation}%
where $y_{i}=x_{i}-\cos \left( x_{i}\right) \sin \left( x_{i}\right) $, $%
x_{i}$ is the solution of Eq.~(\ref{transc}), and
\begin{equation}
Si(x)=\int\limits_{0}^{x}\frac{\sin (t)}{t}dt.  \label{Si}
\end{equation}

Note that the functional of the running coupling $\alpha _{s}(R)$ in Eq. (%
\ref{Cij}) and other parameters (the quark mass $m_{i}$, zero-point energy
coefficient $Z_{0}$, bag constant $B$) are evaluated in Ref. \cite%
{Zhang:2021yul} via mapping the model mass prediction to the ground-state
masses of the observed mesons and baryons. The obtained values for these
model parameters are\cite{Zhang:2021yul}
\begin{equation}
\begin{Bmatrix}
m_{n}=0\,\text{GeV,} & m_{s}=0.279\,\text{GeV,} \\
m_{c}=1.641\,\text{GeV,} & m_{b}=5.093\,\text{GeV,} \\
Z_{0}=1.84, & B^{1/4}=0.145\,\text{GeV.}%
\end{Bmatrix}
\label{originalparas}
\end{equation}%
We will use these parameters to analyze the heavy tetra-
quarks in this work,
with the bag radius $R$ determined variationally via the MIT bag model.

The binding energy $M_{B}$ in Eq. (\ref{dM}) measures the short-range
chromoelectric interaction between quarks and/or antiquarks. For the massive
quarks of $i$ and $j$, this energy, which scales like $-\alpha
_{s}(r_{ij})/r_{ij}$, becomes sizable when both quarks($i$ and $j$) are
massive, moving nonrelativistically. We treat this energy as the sum of the
pair binding energies, $B_{QQ^{\prime }}(B_{Qs})$, between heavy quarks ($Q$
and $Q^{\prime }$) and between heavy quarks $Q$ and the strange quarks $s$%
\cite{Karliner:2014gca,Karliner:2017elp}. This leads to five binding
energies $B_{cs}$, $B_{cc}$, $B_{bs}$, $B_{bb}$, and $B_{bc}$ for any quark
pair in the color configuration $\bar{3}_{c}$, which are extractable from
heavy mesons and can be scaled to other color configurations.

Assuming two quarks $QQ^{\prime }$ to be in the color anti-triplet $\bar{3}%
_{c}$ inside baryon, the binding energy $B_{QQ^{\prime }}\equiv
B_{QQ^{\prime }}[\bar{3}_{c}]$ are extracted in the MIT bag model\cite%
{Zhang:2021yul} (appendix A) for the combination of $QQ^{\prime }=cc$, $bb$,
$bc$, $bs\ $and $cs$, so that a unified parameter setup was established for
the ground states of meson, baryons and heavy hadrons (including doubly
baryon and tetraquar-
ks). The results are\cite{Zhang:2021yul}
\begin{equation}
\begin{Bmatrix}
B_{cs}=-0.025\,\text{GeV,} & B_{cc}=-0.077\,\text{GeV,} \\
B_{bs}=-0.032\,\text{GeV,} & B_{bb}=-0.128\,\text{GeV,} \\
B_{bc}=-0.101\,\text{GeV.} &
\end{Bmatrix}
\label{Bcs}
\end{equation}

\section{Color and spin factors for tetraquarks}

To determine the mass splitting $M_{CMI}=\langle H_{CMI}\rangle $ via the
CMI Hamiltonian $H_{CMI}$ in Eq.~(\ref{CMI}), one has to evaluate the
chromomagnetic matrices $H_{CMI}$ of the tetraquarks $T$ for a given quantum
number $J^{PC}$. For this, one can firstly work out the color factors ${%
\left\langle \boldsymbol{\lambda }_{i}\cdot \boldsymbol{\lambda }%
_{j}\right\rangle }$ and spin factors ${\left\langle \boldsymbol{\sigma }%
_{i}\cdot \boldsymbol{\sigma }_{j}\right\rangle }$ as matrices over the
color and spin bases, respectively,the allowed states of tetraquarks with
given $J^{PC}$ in Table 1. In this section, we present the color and spin
factors as a matrix elements in the color and spin space, and give an
unified expressions for binding energy $M_{B}=\sum_{i<j}F_{c}B_{ij}(\bar{3}%
_{c})$ for the both color :

Color factor in the color states $|n\rangle $ and $|m\rangle $:
\begin{equation}
{\left\langle \boldsymbol{\lambda }_{i}\cdot \boldsymbol{\lambda }%
_{j}\right\rangle }_{nm}=\sum_{\alpha =1}^{8}Tr\left( c_{in}^{\dagger
}\lambda ^{\alpha }c_{im}\right) Tr\left( c_{jn}^{\dagger }\lambda ^{\alpha
}c_{jm}\right) ,  \label{colorfc}
\end{equation}%
and spin factor in the spin states $|x\rangle $ and $|y\rangle $:
\begin{equation}
{\left\langle \boldsymbol{\sigma }_{i}\cdot \boldsymbol{\sigma }%
_{j}\right\rangle }_{xy}=\sum_{\alpha =1}^{3}Tr\left( \chi _{ix}^{\dagger
}\sigma ^{\alpha }\chi _{iy}\right) Tr\left( \chi _{jx}^{\dagger }\sigma
^{\alpha }\chi _{jy}\right) ,  \label{spinfc}
\end{equation}%
where $c_{in}$ stands for color basis (three colors $r$, $g$, and $b$) of a
given quark $i$, and $\chi _{ix}$ represents its spin basis (with two spin
components of $\uparrow $ and $\downarrow $).

In color-spin wavefunction of the tetraquark $T$, one can compute explicitly
the expectation values of $H_{CMI}$,
\begin{equation}
{\left\langle T|H_{CMI}|T\right\rangle =}-\sum_{i<j}{\left\langle
\boldsymbol{\lambda }_{i}\cdot \boldsymbol{\lambda }_{j}\right\rangle }_{TT}{%
\left\langle \boldsymbol{\sigma }_{i}\cdot \boldsymbol{\sigma }%
_{j}\right\rangle }_{TT}C_{ij}\text{,}  \label{EVT}
\end{equation}%
to obtain the color and spin factor, writing the mass formula for $M_{CMI}$
in terms of the CMI couplings \thinspace $C_{ij}$, which are given further
by Eq. (\ref{Cij}) in the MIT bag model. Here the state of $T$ are the mixed
states listed in Table 1, with the mixed weight $w=(w_{1},w_{2},\cdots
,w_{f})$ solved (as eigenvector during the CMI diagonalization) numerically
in Table 2,5-7 in the section 5.

Given the two formula (\ref{colorfc}) and (\ref{spinfc}), one can compute
the color factors ${\left\langle \phi _{1}^{T},\phi _{2}^{T}|\boldsymbol{%
\lambda }_{i}\cdot \boldsymbol{\lambda }_{j}|\phi _{1}^{T},\phi
_{2}^{T}\right\rangle }$ as $2$ by $2$ matrix in the color subspace of $%
\left( \phi _{1}^{T},\phi _{2}^{T}\right) $, via applying Eqs. (\ref{ph1})
and (\ref{ph2}) in appendix A. The result are obtained to be
\begin{equation}
\begin{aligned} &\left\langle \boldsymbol{\lambda_{1}}\cdot
\boldsymbol{\lambda_{2}} \right\rangle= \left\langle
\boldsymbol{\lambda_{3}}\cdot \boldsymbol{\lambda_{4}} \right\rangle=
\begin{bmatrix}\setlength{\arraycolsep}{8pt} \frac{4}{3} & 0 \\ 0 &
-\frac{8}{3} \end{bmatrix},\\ &\left\langle \boldsymbol{\lambda_{1}}\cdot
\boldsymbol{\lambda_{3}} \right\rangle= \left\langle
\boldsymbol{\lambda_{2}}\cdot \boldsymbol{\lambda_{4}} \right\rangle=
\begin{bmatrix} -\frac{10}{3} & 2\sqrt{2} \\ 2\sqrt{2} & -\frac{4}{3}
\end{bmatrix}, \\ &\left\langle \boldsymbol{\lambda_{1}}\cdot
\boldsymbol{\lambda_{4}} \right\rangle= \left\langle
\boldsymbol{\lambda_{2}}\cdot \boldsymbol{\lambda_{3}} \right\rangle=
\begin{bmatrix} -\frac{10}{3} & -2\sqrt{2} \\ -2\sqrt{2} & -\frac{4}{3}
\end{bmatrix}. \end{aligned}  \label{cfcT}
\end{equation}%
From the above matrices, we see that the color conifgurations $\phi
_{1}^{T}\ $and $\phi _{2}^{T}$ may mix for a tetraquark state $T$ due to the
chromomagnetic interaction.

We further consider the binding energy $M_{B}$ based on Eq. (\ref{Bcs}),
which corresponds to the binding energy $B_{ij}\equiv B_{ij}[\bar{3}_{c}]$
in baryons with the quark pair ($i,j$) in $\bar{3}_{c}$. Let us then
consider the binding energy $M_{B}$ for a given color configurations of the
tetraquark $T=(q_{1}q_{2})^{R}(\bar{q}_{3}\bar{q}_{4})^{\bar{R}}$(with
representation $R=$ $6_{c}$ and $\bar{3}_{c}$). First of all, one can scale
the pair binding energy $B_{ij}\equiv B_{ij}[\bar{3}_{c}]$ of the pair in
baryon to $F_{c}[R]B_{ij}[\bar{3}_{c}]$ of the pair in tetraquark $T$, where
$F_{c}[R]$ is the ratio of the color factor in Eq. (\ref{cfcT}) to the color
factor ${\left\langle \boldsymbol{\lambda }_{i}\cdot \boldsymbol{\lambda }%
_{j}\right\rangle }_{B}=-8/3$ for baryon with each of quark pair ($i,j$) in $%
\bar{3}_{c}$. At last, applying to all quark pair ($i,j$) of the tetraquark $%
T$ with configurations $\phi _{1}^{T}\ $and $\phi _{2}^{T}$, one can obtain
the pair binding energies $F_{c}[R]B_{ij}$, whose sums are,
\begin{equation}
M_{B}(\phi _{1}^{T})=-\frac{1}{2}B_{12}-\frac{1}{2}B_{34}+\frac{5}{4}B_{13}+%
\frac{5}{4}B_{14}+\frac{5}{4}B_{23}+\frac{5}{4}B_{24},  \label{TB1}
\end{equation}%
\begin{equation}
M_{B}(\phi _{2}^{T})=B_{12}+B_{34}+\frac{1}{2}B_{13}+\frac{1}{2}B_{14}+\frac{%
1}{2}B_{23}+\frac{1}{2}B_{24},  \label{TB2}
\end{equation}%
for the tetraquark $T$, respectively, where $B_{ij}$ is the binding energy
with ($i,j$) in $\bar{3}_{c}$.

For color sextets of the pair ($1,2$) and ($3,4$), for instance, the binding
energy is $-B_{12}/2$ and $-B_{34}/2$, respectively, with $%
F_{c}=(4/3)/(-8/3)=-1/2$. For any of representation of the quark $i$ and
antiquark $j$, the binding energies in $T$ are either $-5B_{ij}/4$ or $%
B_{ij}/2$. We note that $B_{ij}$ vanishes if both of quark $i$ and $j$ are
light quarks or one of them is non-strange light quark ($B_{nQ}=0$, $%
B_{nn}=0 $, $B_{n\bar{n}}=0$, $B_{s\bar{s}}=0$) since the short range
interactions between ($i,j$) quarks are small and thereby ignorable
averagely for quark pair $(i,j)=(n,n)$ or $(i,j)=(s,s)$, due to their
relativistic motion.

We come to consider the spin factors, which is given by ${\left\langle \chi
^{T}|\boldsymbol{\sigma }_{i}\cdot \boldsymbol{\sigma }_{j}|\chi
^{T}\right\rangle }$. In the subspace spanned by $\{\chi _{1\sim 6}^{T}\}$
in Eq. (\ref{pspinT}), the direct computation yields the following matrices,
\begin{equation}
\setlength{\arraycolsep}{8pt}\left\langle \boldsymbol{\sigma _{1}}\cdot
\boldsymbol{\sigma _{2}}\right\rangle _{\chi _{1}^{T}}=%
\begin{bmatrix}
1 & 0 & 0 & 0 & 0 & 0 \\
0 & 1 & 0 & 0 & 0 & 0 \\
0 & 0 & 1 & 0 & 0 & 0 \\
0 & 0 & 0 & 1 & 0 & 0 \\
0 & 0 & 0 & 0 & -3 & 0 \\
0 & 0 & 0 & 0 & 0 & -3%
\end{bmatrix}%
,  \label{sf1}
\end{equation}%
\begin{equation}
\setlength{\arraycolsep}{2.3pt}\left\langle \boldsymbol{\sigma _{1}}\cdot
\boldsymbol{\sigma _{3}}\right\rangle _{\chi _{2}^{T}}=%
\begin{bmatrix}
1 & 0 & 0 & 0 & 0 & 0 \\
0 & -1 & 0 & \sqrt{2} & -\sqrt{2} & 0 \\
0 & 0 & -2 & 0 & 0 & -\sqrt{3} \\
0 & \sqrt{2} & 0 & 0 & 1 & 0 \\
0 & -\sqrt{2} & 0 & 1 & 0 & 0 \\
0 & 0 & -\sqrt{3} & 0 & 0 & 0%
\end{bmatrix}%
,  \label{sf2}
\end{equation}%
\begin{equation}
\setlength{\arraycolsep}{3pt}\left\langle \boldsymbol{\sigma _{1}}\cdot
\boldsymbol{\sigma _{4}}\right\rangle _{\chi _{3}^{T}}=%
\begin{bmatrix}
1 & 0 & 0 & 0 & 0 & 0 \\
0 & -1 & 0 & -\sqrt{2} & -\sqrt{2} & 0 \\
0 & 0 & -2 & 0 & 0 & \sqrt{3} \\
0 & -\sqrt{2} & 0 & 0 & -1 & 0 \\
0 & -\sqrt{2} & 0 & -1 & 0 & 0 \\
0 & 0 & \sqrt{3} & 0 & 0 & 0%
\end{bmatrix}%
,  \label{sf3}
\end{equation}%
\begin{equation}
\setlength{\arraycolsep}{5.5pt}\left\langle \boldsymbol{\sigma _{2}}\cdot
\boldsymbol{\sigma _{3}}\right\rangle _{\chi _{4}^{T}}=%
\begin{bmatrix}
1 & 0 & 0 & 0 & 0 & 0 \\
0 & -1 & 0 & \sqrt{2} & \sqrt{2} & 0 \\
0 & 0 & -2 & 0 & 0 & \sqrt{3} \\
0 & \sqrt{2} & 0 & 0 & -1 & 0 \\
0 & \sqrt{2} & 0 & -1 & 0 & 0 \\
0 & 0 & \sqrt{3} & 0 & 0 & 0%
\end{bmatrix}%
,  \label{sf4}
\end{equation}%
\begin{equation}
\setlength{\arraycolsep}{2.3pt}\left\langle \boldsymbol{\sigma _{2}}\cdot
\boldsymbol{\sigma _{4}}\right\rangle _{\chi _{5}^{T}}=%
\begin{bmatrix}
1 & 0 & 0 & 0 & 0 & 0 \\
0 & -1 & 0 & -\sqrt{2} & \sqrt{2} & 0 \\
0 & 0 & -2 & 0 & 0 & -\sqrt{3} \\
0 & -\sqrt{2} & 0 & 0 & 1 & 0 \\
0 & \sqrt{2} & 0 & 1 & 0 & 0 \\
0 & 0 & -\sqrt{3} & 0 & 0 & 0%
\end{bmatrix}%
,  \label{sf5}
\end{equation}%
\begin{equation}
\setlength{\arraycolsep}{8pt}\left\langle \boldsymbol{\sigma _{3}}\cdot
\boldsymbol{\sigma _{4}}\right\rangle _{\chi _{6}^{T}}=%
\begin{bmatrix}
1 & 0 & 0 & 0 & 0 & 0 \\
0 & 1 & 0 & 0 & 0 & 0 \\
0 & 0 & 1 & 0 & 0 & 0 \\
0 & 0 & 0 & -3 & 0 & 0 \\
0 & 0 & 0 & 0 & 1 & 0 \\
0 & 0 & 0 & 0 & 0 & -3%
\end{bmatrix}%
.  \label{sf6}
\end{equation}

Combing the spin factors in Eqs. (\ref{sf1})-(\ref{sf6}) with Eqs. (\ref%
{cfcT}), we are the position to use Eqs. (\ref{EVT}), Eq. (\ref{colorfc})
and Eq. (\ref{spinfc}) to compute the mass splitting $M_{CMI}$ duo to
chromomagnetic interaction. Using Eqs. (\ref{TB1}) and (\ref{TB2}), one can
compute the mass sum $\Delta M=M_{B}+M_{CMI\text{ }}$in Eq. \ref{dM} and
further obtain, via adding mass of the bag $M_{bag}=\sum_{i}\omega
_{i}+(4/3)\pi R^{3}B-Z_{0}/R$, a complete mass formula for the hidden
heavy-flavor tetraquark systems $T$ addressed in this work,
\begin{equation}
M(T)=M_{bag}+M_{B}+M_{CMI\text{ }}(C_{ij}),  \label{MT}
\end{equation}%
in which $M_{CMI\text{ }}(C_{ij})$ are linear functions of the CMI couplings
\thinspace $C_{ij}$, with the linear coefficients given by the color and
spin factors shown in this section.

\section{Masses of hidden heavy-flavor tetraquarks}

Given the input parameters in Eqs. (\ref{originalparas}), one can
numerically solve Eq. (\ref{MBm}) variationally, with the mass splitting $%
M_{CMI}$ and the CMI couplings \thinspace $C_{ij}$ given by Eqs. (\ref{Cij}%
), (\ref{alphaS-mine}), (\ref{muBari}) and (\ref{Iij}), to obtain bag radius
$R$ and numerically give the masses $M(T)$ of the hidden heavy-flavor
tetraquarks $T$. Meanwhile, we show the numerical corresponding results for
the bag radius $R_{0}$, the mixing weights (eigenvectors of the CMI matrix $%
H_{CMI\text{ }}$), the tetraquark masses $M(T)$ and thresholds of two mesons
as a final states in the Tables 2, 5-7. In the following, we present the
results and discussions with respect to the tetraquark systems addressed
below in order.

\subsection{Fully heavy tetraquark systems}

\begin{figure}[th]
\centering
\includegraphics[width=0.5\textwidth]{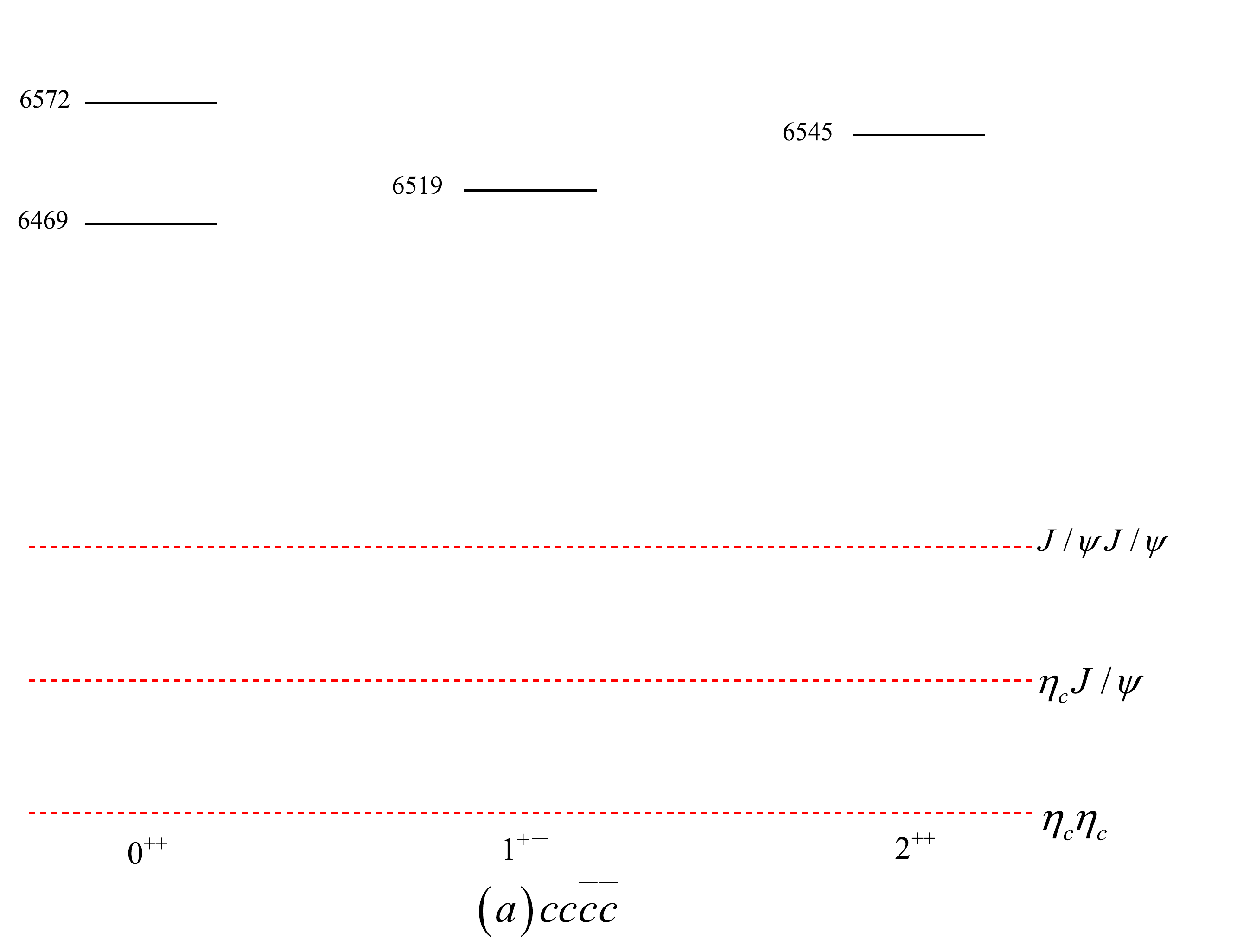}
\caption{The computed masses (MeV the solid lines) of the $cc\bar{c}\bar{c}$
tetraquark system in their ground-states, as well as two meson thresholds
(MeV the dotted lines). }
\label{a}
\end{figure}

In the case of fully charmed systems of the tetraquarks $cc\bar{c}\bar{c}$,
we show the numerical results for $R_{0}$, the state-mixing weights
(eigenvectors of $H_{CMI\text{ }}$), the tetraquark masses $M(T)$ and
thresholds of two mesons final states in the Table \ref{tab:ccccbbbbmass},
with the later two plotted in Fig. \ref{a}. We see that for $J^{PC}=0^{++}$
there are two states of the tetraquarks $cc\bar{c}\bar{c}$ with the masses
of $6572\,$MeV and $6469\,$MeV, splitted by $103\,$MeV. The tetraquark ($cc%
\bar{c}\bar{c}$) states with $J^{PC}=1^{+-}$ and $J^{PC}=2^{++}$ have the
masses within a similar mass region, as shown in Fig \ref{a}. We find that
all these $cc\bar{c}\bar{c}$ states relatively far above their two mesons
thresholds shown explicitly. For instance, the $0^{++}$ state are all above
the thresholds of the $J/\psi J/\psi $ and $\eta _{c}\eta _{c}$, about $%
275-605\,$MeV, indicating that they are not stable against strong decays
through quark rearrangement to the final state of $J/\psi J/\psi $ as well
as $\eta _{c}\eta _{c}$. For the $1^{+-}$ state, there is one state, and its
mass is above the thresholds of the two mesons $\eta _{c}J/\psi $ and $%
J/\psi J/\psi $ about $325-440\,$MeV, unstable against the strong decay to
the later. In the case of the $2^{++}$ state, there is one state with the
mass above the threshold($J/\psi$$J/\psi $) about $350\,$MeV, also strongly unstable. We also compare our
calculations with other works cited and list the results in Table \ref%
{tab:cccc}.

\begin{figure}[th]
\centering
\includegraphics[width=0.5\textwidth]{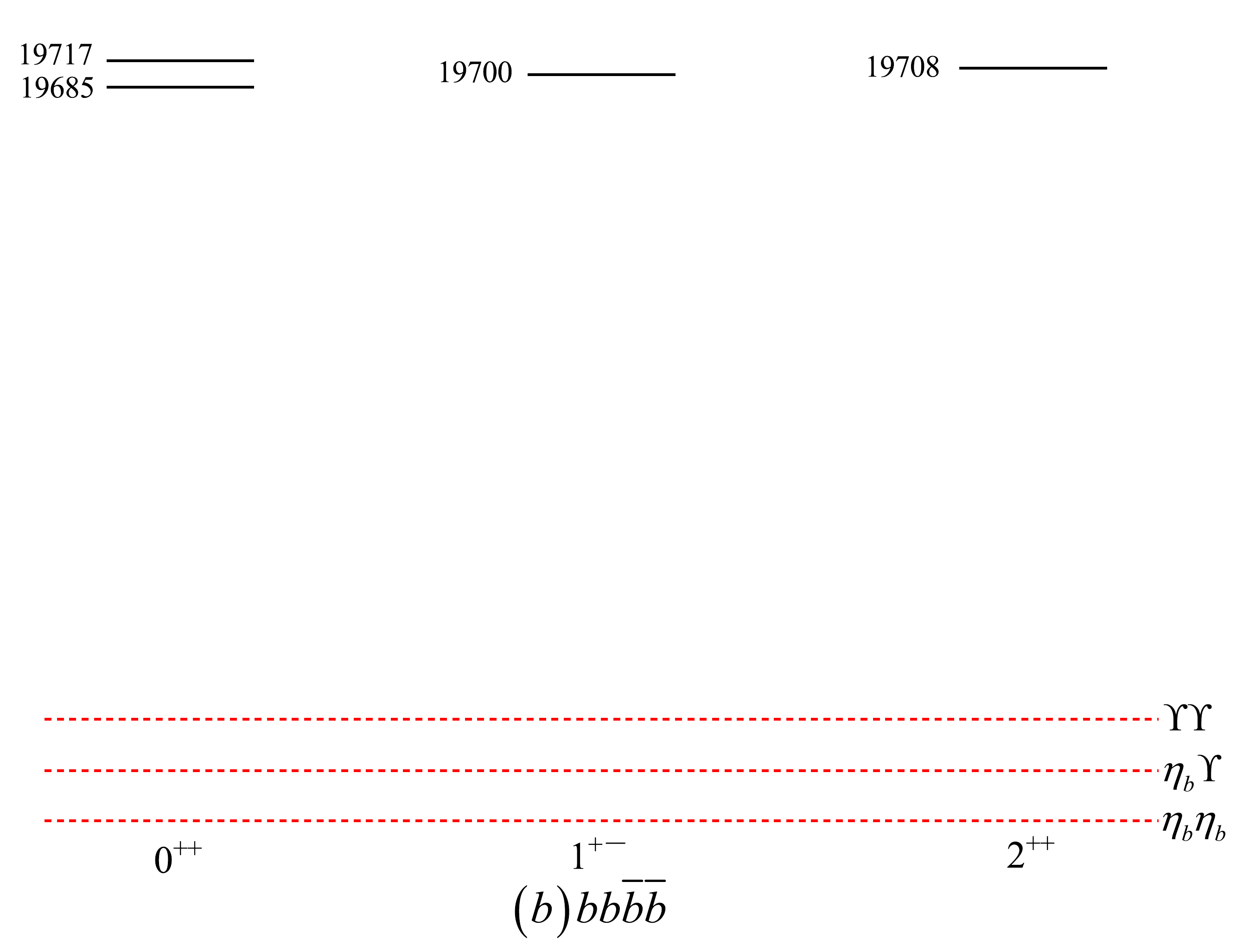}
\caption{The computed masses (MeV the solid lines) of the $bb\bar{b}\bar{b}$
tetraquark system in their ground-states, as well as two meson thresholds
(MeV the dotted lines). }
\label{b}
\end{figure}

\renewcommand{\tabcolsep}{0.5cm} \renewcommand{\arraystretch}{2.05}
\begin{table*}[tbh]
\caption{The numerical results for the bag radius $R_{0}$, the state-mixing
weights (eigenvectors of $H_{CMI}$), the tetraquark masses $M(T)$ and
thresholds of two mesons final states for the hidden heavy-flavor
tetraquarks($cc\bar{c}\bar{c}$, $bb\bar{b}\bar{b}$)}
\label{tab:ccccbbbbmass}%
\begin{tabular}{ccccccc}
\hline\hline
\textrm{State} & $J^{PC}$ & \textrm{Eigenvector} & $R_{0}$(GeV$^{-1}$) & $%
M(T)$(MeV) & Threshold (MeV) &  \\ \hline
$cc\bar{c}\bar{c}$ & $0^{++}$ & $(0.58,0.81)$ & $4.74$ & $6572$ & $J/\psi
J/\psi =6194$;$\eta _{c}\eta _{c}=5967$ &  \\
&  & $(-0.81,0.58)$ & $4.44$ & $6469$ &  &  \\
& $1^{+-}$ & $1.00$ & $4.59$ & $6519$ & $\eta _{c}J/\psi =6080$;$J/\psi
J/\psi =6194$ &  \\
& $2^{++}$ & $1.00$ & $4.66$ & $6545$ & $J/\psi J/\psi =6194$ &  \\
$bb\bar{b}\bar{b}$ & $0^{++}$ & $(0.58,0.81)$ & $3.15$ & $19717$ & $\Upsilon
\Upsilon =18921$;$\eta _{b}\eta _{b}=18798$ &  \\
&  & $(-0.81,0.58)$ & $2.99$ & $19685$ &  &  \\
& {$1^{+-}$} & $1.00$ & $3.07$ & $19700$ & $\eta _{b}\Upsilon =18859$;$%
\Upsilon \Upsilon =18921$ &  \\
& $2^{++}$ & $1.00$ & $3.11$ & $19708$ & $\Upsilon \Upsilon =18921$ &  \\
\hline\hline
\end{tabular}%
\end{table*}

\renewcommand{\tabcolsep}{0.34cm} \renewcommand{\arraystretch}{2.3}
\begin{table*}[tbh]
\caption{Comparision of our results for the $cc\bar{c}\bar{c}$ systems with
other calculations cited.All masses are in unit of MeV.}
\label{tab:cccc}%
\begin{tabular}{ccccccccccc}
\hline\hline
\textrm{State} & $J^{PC}$ & \textrm{This\ work} & \cite{Liu:2019zuc} & \cite%
{Lloyd:2003yc} & \cite{Wu:2016vtq} & \cite{Chen:2016jxd} & \cite{Ader:1981db}
& \cite{Iwasaki:1975pv} & \cite{Karliner:2016zzc} & \cite{Barnea:2006sd} \\
\hline
${(cc\bar{c}\bar{c})}$ & $0^{++}$ & $6469$ & $6487$ & $6477$ & $6797$ & $%
6440-6820$ & $6437$ & $6200$ & $6192$ & $6038-6115$ \\
& $0^{++}$ & $6572$ & $6518$ & $6695$ & $7016$ & $6440-6820$ & $6383$ & $...$
& $...$ & $...$ \\
& $1^{+-}$ & $6519$ & $6500$ & $6528$ & $6899$ & $6370-6510$ & $6437$ & $...$
& $...$ & $6101-6176$ \\
& $2^{++}$ & $6545$ & $6524$ & $6573$ & $6956$ & $6370-6510$ & $6437$ & $...$
& $...$ & $6172-6216$ \\ \hline\hline
\end{tabular}%
\end{table*}

For fully bottom systems of the tetraquarks $bb\bar{b}\bar{b}$, the solved
results of the model are shown in Table \ref{tab:ccccbbbbmass}. We find that
all these $bb\bar{b}\bar{b}$ states (with $J^{PC}=0^{++}$, $1^{+-}$ and $%
2^{++}$ ) are close to each other and strongly unstable as they are far
above their two mesons final states shown. For instance, two of the $0^{++}$
states have the masses of $19717\,$MeV and $19685\,$MeV (with mass splitting
$32\,$MeV). As seen in Fig \ref{b}, the two of the $0^{++}$ states are above
thresholds ($\Upsilon \Upsilon $, $\eta _{b}\eta_{b} $) about $764-919\,$%
MeV. For the $1^{+-}$ state of $bb\bar{b}\bar{b}$, its mass is higher than
the threshold($\eta _{b}\Upsilon $ and $\Upsilon \Upsilon$) about $780-840\,$%
MeV. For the $2^{++}$ state, it is above the threshold ($\Upsilon \Upsilon$%
), about $787\,$MeV. By the way, our results for the $bb\bar{b}\bar{b}$
systems are also compared to other works cited, as shown in Table \ref%
{tab:bbbb}. \renewcommand{\tabcolsep}{0.66cm} \renewcommand{%
\arraystretch}{2.3}
\begin{table*}[tbh]
\caption{Comparision of our results for the $bb\bar{b}\bar{b}$ systems with
other calculations cited.All masses are in unit of MeV. }
\label{tab:bbbb}%
\begin{tabular}{cccccccc}
\hline\hline
\textrm{State} & $J^{PC}$ & \textrm{This\ work} & \cite{Liu:2019zuc} & \cite%
{Wu:2016vtq} & \cite{Wang:2017jtz,Wang:2018poa} & \cite{Karliner:2016zzc} &
\cite{Berezhnoy:2011xn} \\ \hline
${(bb\bar{b}\bar{b})}$ & $0^{++}$ & 19685 & 19322 & 20155 & 18840 & 18826 &
18754 \\
& $0^{++}$ & 19717 & 19338 & 20275 & ... & ... & ... \\
& $1^{+-}$ & 19700 & 19329 & 20212 & 18840 & ... & 18808 \\
& $2^{++}$ & 19708 & 19341 & 20243 & 18850 & ... & 18916 \\ \hline\hline
\end{tabular}%
\end{table*}

\subsection{The bottom-charmed system($cb\bar{c}\bar{b}$)}

\begin{figure}[th]
\centering
\includegraphics[width=0.5\textwidth]{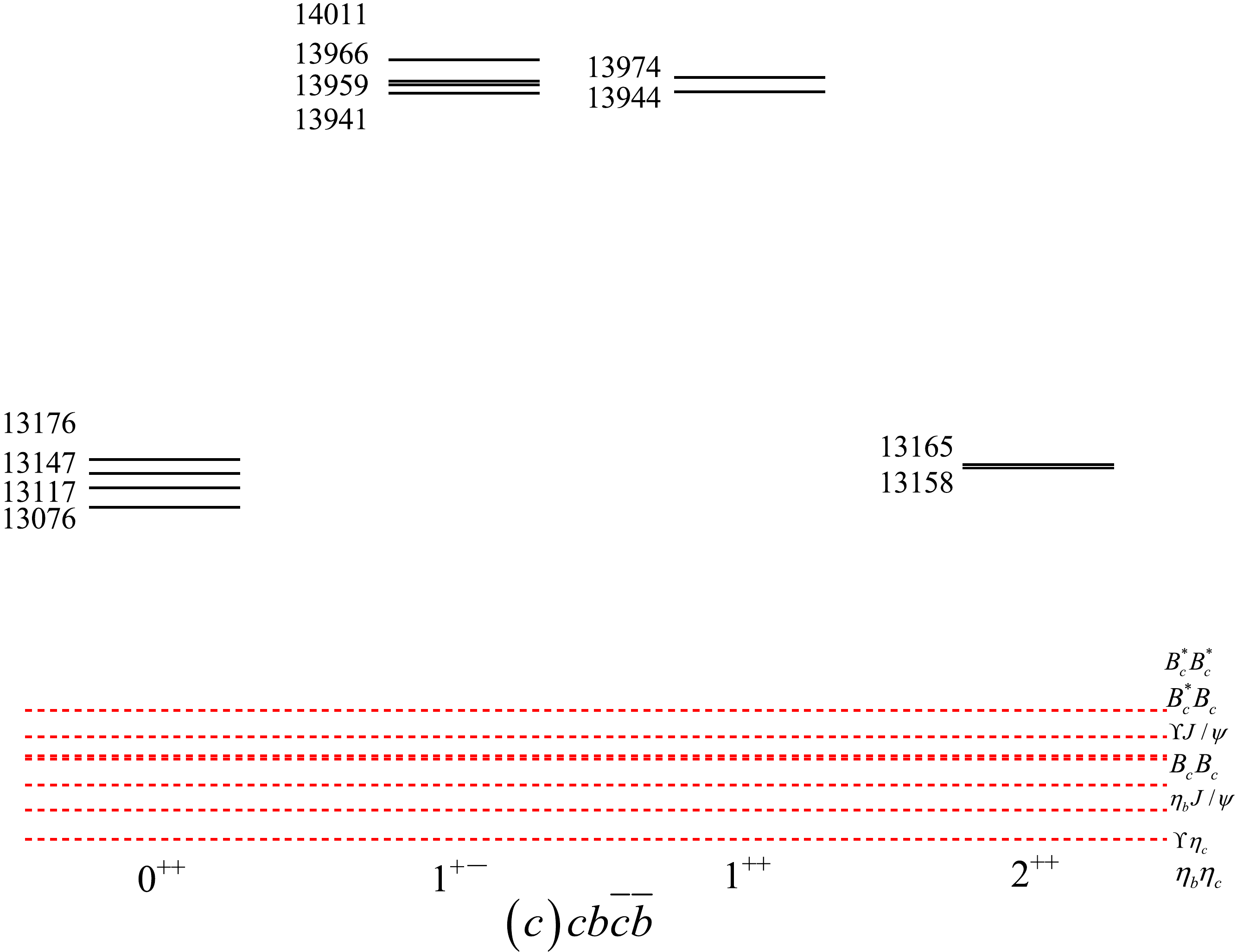}
\caption{Computed masses (MeV the solid lines) of the $cb\bar{c}\bar{b}$ systems
of tetraquarks in their ground states, and the thresholds (MeV the dotted lines)
of the two meson final states. }
\label{cb}
\end{figure}

For bottom-charmed systems of the tetraquarks $cb\bar{c}\bar{b}$, we show in
Table \ref{tab:cbcbmass} the computed results for $R_{0}$, the mixing
weights (the CMI eigenvectors), the tetraquark masses $M(T)$ and thresholds
(two mesons), with the later two plotted in Fig \ref{cb}. We find that there
are four $J^{PC}=0^{++}$ states for the $cb\bar{c}\bar{b}$ systems, all
above the thresholds ($B_{c}^{\ast }B_{c}^{\ast }$, $\Upsilon J/\psi $, $%
B_{c}B_{c}$ and $\eta _{b}\eta _{c}$) about $424-794\,$MeV. There are four
states of the $cb\bar{c}\bar{b}$ systems with $J^{PC}=1^{+-}$, all highly
above the thresholds ($B_{c}^{\ast }B_{c}^{\ast }$, $B_{c}^{\ast }B_{c}$, $%
\eta _{b}J/\psi $ and $\Upsilon \eta _{c}$) about $1289-1567\,$MeV, and two
states of the $cb\bar{c}\bar{b}$ systems with $J^{PC}=1^{++}$, all highly
above the thresholds ($B_{c}^{\ast }B_{c}$ and $\Upsilon J/\psi $) about $%
1347-1417\,$MeV. There are also two states with $J^{PC}=2^{++}$, both above
the thresholds ($B_{c}^{\ast }B_{c}^{\ast }$ and $\Upsilon J/\psi $) about $%
506-608\,$MeV. This indicates that the $cb\bar{c}\bar{b}$ systems are
unstable against strong decay to the final states of the mesons.

\renewcommand{\tabcolsep}{0.41cm} \renewcommand{\arraystretch}{2.05}
\begin{table*}[!htb]
\caption{Computed results for the bottom-charmed tetraquark states $cb%
\bar{c}\bar{b}$. The thresholds of two mesons are also
listed.}
\label{tab:cbcbmass}%
\begin{tabular}{ccccccc}
\hline\hline
\textrm{State} & $J^{PC}$ & \textrm{Eigenvector} & $R_{0}$(GeV$^{-1}$) & $%
M(T)$(MeV) & \textrm{Threshold} (MeV) &  \\ \hline
$cb\bar{c}\bar{b}$ & $0^{++}$ & (-0.21,-0.52,0.82,0.16) & 3.76 & 13076 & $%
B_{c}^{*}B_{c}^{*}=12652$;$\Upsilon J/\psi=12557$ &  \\
&  & (-0.77, 0.24,-0.16,0.58) & 3.90 & 13117 & $B_{c}B_{c}=12550$;$\eta _{b}
\eta _{c}=12382$ &  \\
&  & (-0.14,-0.82,-0.55,0.01) & 4.02 & 13147 &  &  \\
&  & (0.59,-0.06,-0.05,0.80) & 4.12 & 13176 &  &  \\
& $1^{+-}$ & (-0.43,0.39,0.14,0.80) & 3.95 & 13941 & $%
B_{c}^{*}B_{c}^{*}=12652$;$B_{c}^{*}B_{c}=12597$ &  \\
&  & (0.63,0.74,-0.25,0.02) & 4.0 & 13959 & $\eta _{b}J/\psi=12496$;$%
\Upsilon \eta _{c}=12444$ &  \\
&  & (0.57,-0.26,0.71,0.32) & 4.04 & 13966 &  &  \\
&  & (0.30,-0.48,-0.65,0.51) & 4.18 & 14011 &  &  \\
& $1^{++}$ & (-0.58,0.82) & 3.94 & 13944 & $B_{c}^{*}B_{c}=12597$;$\Upsilon
J/\psi=12557$ &  \\
&  & (0.82,0.58) & 4.07 & 13974 &  &  \\
& $2^{++}$ & (0.74,0.67) & 4.06 & 13158 & $B_{c}^{*}B_{c}^{*}=12652$;$%
\Upsilon J/\psi=12557$ &  \\
&  & (-0.68,0.74) & 4.08 & 13165 &  &  \\ \hline\hline
\end{tabular}%
\end{table*}

\subsection{The strange-heavy systems ($sc\bar{s}\bar{c}$ and $sb\bar{s}\bar{%
b}$)}

\begin{figure}[th]
\centering
\includegraphics[width=0.5\textwidth]{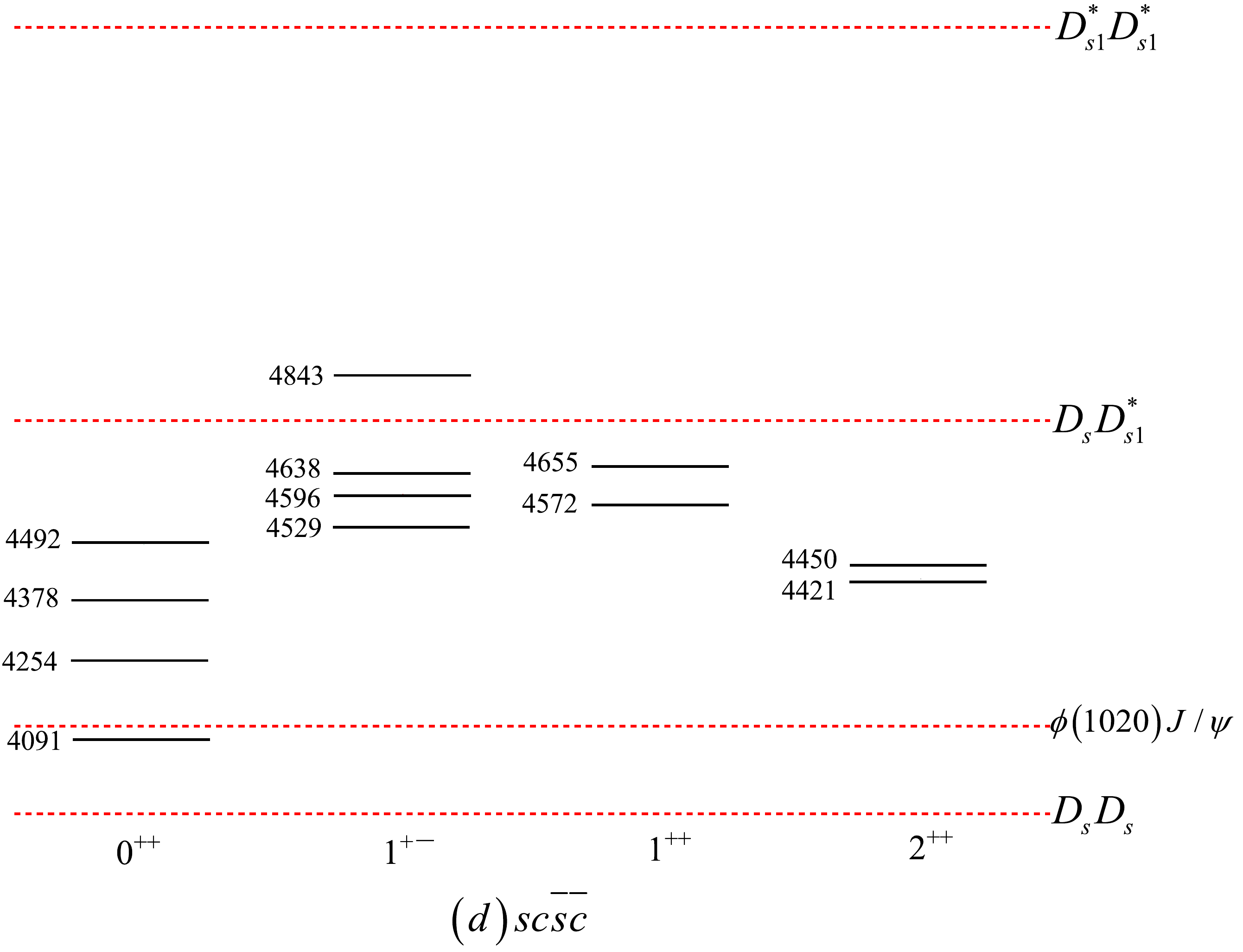}
\caption{Computed masses (Mev the solid lines) of the $sc\bar{s}\bar{c}$
tetraquarks and corresponding two meson thresholds (MeV the dotted lines)}
\label{sc}
\end{figure}
\begin{figure}[th]
\centering
\includegraphics[width=0.5\textwidth]{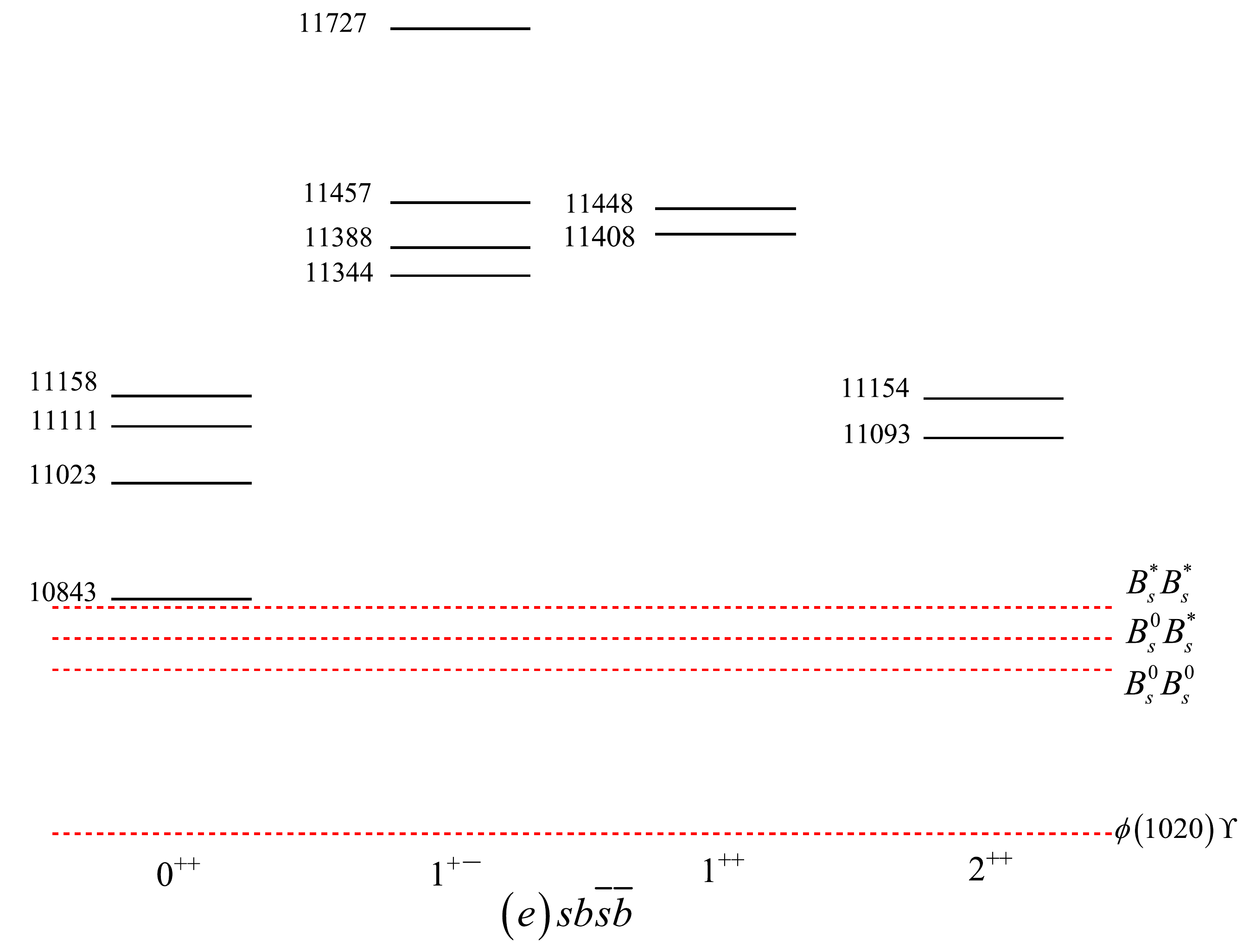}
\caption{Computed masses (MeV the solid lines) of the $sb\bar{s}\bar{b}$
tetraquarks and corresponding two meson thresholds (MeV the dotted lines)}
\label{sb}
\end{figure}

For strange-charmed systems of the tetraquarks $sc\bar{s}\bar{c}$, we show
in Table \ref{tab:scscsbsbmass} the computed results for $R_{0}$, the mixing
weights, the masses $M(T)$ and thresholds (two mesons), with the later two
plotted in Fig \ref{sc}. We find that there are four $J^{PC}=0^{++}$ states
of the $sc\bar{s}\bar{c}$ systems, all below the threshold of $D_{s1}^{\ast
}D_{s1}^{\ast }$, in which three states with masses ($4492,4378,4254$)$\,$%
MeV are above the thresholds of $D_{s}D_{s}$ and $\phi \left( 1020\right)
J/\psi $ about $137-556\,$
MeV and unstable against strong decay to them. The
lowest state with mass of $4091\,$MeV is above the threshold of $D_{s}D_{s}$
about $155\,$MeV while it is near to the threshold of $\phi (1020)J/\psi $,
far below the threshold of $D_{s1}^{\ast }D_{s1}^{\ast }$. It is uncertain
whether the lowest state is above or below the threshold of $\phi \left(
1020\right) J/\psi $ as the model uncertainty is as large as $\pm 40\,$MeV%
\cite{Zhang:2021yul}. In the case of the $J^{PC}=1^{+-}$ states, there are
four states, with three of them having the mass of ($4529,4596,4638$)$\,$MeV
and all all below the thresholds of $D_{s}D_{s1}^{\ast }$ and $D_{s1}^{\ast
}D_{s1}^{\ast }$ about 110-1031$\,$MeV and one state, with mass of $4843\,$%
MeV, above the threshold of $D_{s}D_{s1}^{\ast }$ about $95\,$MeV but below
the threshold of $D_{s1}^{\ast }D_{s1}^{\ast }$ about $717\,$MeV. There are
two $sc\bar{s}\bar{c}$ systems with $J^{PC}=1^{++}$, both of which are above
the threshold of $\phi \left( 1020\right) J/\psi $ about $455-538\,$MeV and
below the threshold of $D_{s}D_{s1}^{\ast }$ about $93-176\,$MeV. There are
two $sc\bar{s}\bar{c}$ systems with $J^{PC}=2^{++}$, both above the
threshold of $\phi \left( 1020\right) J/\psi $ about $304-333\,$MeV and
below the threshold of $D_{s1}^{\ast }D_{s1}^{\ast }$ about $1110-1139\,$%
MeV, unstable to strong decay to $\phi \left( 1020\right) J/\psi $.

For strange-bottom systems $sb\bar{s}\bar{b}$, we show in Table \ref%
{tab:scscsbsbmass} the computed results for $R_{0}$, the mixing weights, the
masses $M(T)$ and thresholds, with the later two plotted in Fig \ref{sb}.
Similarly, there are four states for each of $J^{PC}=0^{++}$ and $%
J^{PC}=1^{+-}$, and two states for each of $J^{PC}=1^{++}$ and $J^{PC}=2^{++}
$. All of the $sb\bar{s}\bar{b}$ systems are above the thresholds except for
the lowest one with mass of $10843$ MeV which is near to thresholds ($13\,$%
MeV) of the $B_{s}^{\ast }B_{s}^{\ast }$. The $0^{++}$ states of the $sb\bar{%
s}\bar{b}$ systems are above the thresholds of $B_{s}^{0}B_{s}^{0}$, $%
B_{s}^{\ast }B_{s}^{\ast }$ and $\phi \left( 1020\right) \Upsilon $. Among
them, the minimum mass of $10843\,$MeV can be strongly decayed into $%
B_{s}^{0}B_{s}^{0}$ and $\phi \left( 1020\right) \Upsilon $. Because of the
error in the model, it is uncertain whether it is above or below the
threshold of $B_{s}^{\ast }B_{s}^{\ast }$. The four $J^{PC}=1^{+-}$ states
are all highly above the thresholds of $B_{s}^{0}B_{s}^{\ast }$, $%
B_{s}^{\ast }B_{s}^{\ast }$ (about $562-945\,$MeV and $514-897\,$MeV,
respectively).There are two states $J^{PC}=1^{++}$, which are higher than $%
B_{s}^{0}B_{s}^{\ast }$ and $\phi \left( 1020\right) \Upsilon $(about $626-666\,$MeV and $928-968\,$MeV, respectively). $J^{PC}=2^{++}$
has two states, which are higher than $B_{s}^{\ast }B_{s}^{\ast }$ and $\phi
\left( 1020\right) \Upsilon $ thresholds (about $263-324\,$MeV and $613-674\,$MeV), indicating they are unstable.

\renewcommand{\tabcolsep}{0.315cm} \renewcommand{\arraystretch}{2.05}
\begin{table*}[tbh]
\caption{Computed results for the strange-heavy tetraquark states $sc\bar{s}%
\bar{c}$ and $sb\bar{s}\bar{b}$. The thresholds of two mesons are also
listed. }
\label{tab:scscsbsbmass}%
\begin{tabular}{ccccccc}
\hline\hline
\textrm{State} & $J^{PC}$ & \textrm{Eigenvector} & $R_{0}$(GeV$^{-1}$) & $%
M(T)$(MeV) & \textrm{Threshold} (MeV) &  \\ \hline
$sc\bar{s}\bar{c}$ & $0^{++}$ & $(-0.18,-0.51,0.83,0.14)$ & $4.70$ & $4091$
& $D_{s}D_{s}=3936$;$D_{s1}^{\ast }D_{s1}^{\ast }=5560$ &  \\
&  & $(-0.76,0.21,-0.15,0.60)$ & $4.93$ & $4254$ & $\phi \left( 1020\right)
J/\psi =4117$ &  \\
&  & $(-0.12,-0.84,-0.54,0.01)$ & $5.11$ & $4378$ &  &  \\
&  & $(0.62,-0.06,-0.04,0.78)$ & $5.33$ & $4492$ &  &  \\
& $1^{+-}$ & $(-0.33,0.57,0.03,0.75)$ & $5.21$ & $4529$ & $D_{s}D_{s1}^{\ast
}=4748$;$D_{s1}^{\ast }D_{s1}^{\ast }=5560$ &  \\
&  & $(0.87,0.49,0.08,0.02)$ & $5.30$ & $4596$ &  &  \\
&  & $(0.18,-0.46,0.77,0.41)$ & $5.38$ & $4638$ &  &  \\
&  & $(0.32,-0.47,-0.64,0.52)$ & $5.46$ & $4843$ &  &  \\
& $1^{++}$ & $(-0.58,0.82)$ & $5.22$ & $4572$ & $\phi \left( 1020\right)
J/\psi =4117$;$D_{s}D_{s1}^{\ast }=4748$ &  \\
&  & $(0.82,0.58)$ & $5.41$ & $4655$ &  &  \\
& $2^{++}$ & $(0.55,0.83)$ & $5.39$ & $4421$ & $D_{s1}^{\ast }D_{s1}^{\ast
}=5560$;$\phi \left( 1020\right) J/\psi =4117$ &  \\
&  & $(-0.83,0.55)$ & $5.39$ & $4450$ &  &  \\
$sb\bar{s}\bar{b}$ & $0^{++}$ & $(-0.36,-0.35,0.80,0.33)$ & $4.43$ & $10843$
& $B_{s}^{0}B_{s}^{0}=10734$;$B_{s}^{\ast }B_{s}^{\ast }=10830$ &  \\
&  & $(-0.58,0.39,-0.36,0.62)$ & $4.60$ & $11023$ & $\phi \left( 1020\right)
\Upsilon =10480$ &  \\
&  & $(0.35,0.81,0.47,0.10)$ & $4.74$ & $11111$ &  &  \\
&  & $(0.64,-0.30,-0.13,0.70)$ & $4.86$ & $11158$ &  &  \\
& $1^{+-}$ & $(-0.41,0.67,-0.20,0.58)$ & $4.88$ & $11344$ & $%
B_{s}^{0}B_{s}^{\ast }=10782$;$B_{s}^{\ast }B_{s}^{\ast }=10830$ &  \\
&  & $(0.76,0.39,0.45,0.26)$ & $4.94$ & $11388$ &  &  \\
&  & $(-0.31,-0.47,0.63,0.53)$ & $4.88$ & $11457$ &  &  \\
&  & $(0.38,-0.43,-0.59,0.56)$ & $5.11$ & $11727$ &  &  \\
& $1^{++}$ & $(0.82,0.58)$ & $4.98$ & $11408$ & $B_{s}^{0}B_{s}^{\ast
}=10782 $;$\phi \left( 1020\right) \Upsilon =10480$ &  \\
&  & $(-0.58,0.82)$ & $4.85$ & $11448$ &  &  \\
& $2^{++}$ & $(0.51,0.86)$ & $4.97$ & $11093$ & $B_{s}^{\ast }B_{s}^{\ast
}=10830$;$\phi \left( 1020\right) \Upsilon =10480$ &  \\
&  & $(-0.86,0.51)$ & $4.99$ & $11154$ &  &  \\ \hline\hline
\end{tabular}%
\end{table*}

\subsection{The heavy-light(non-strange) systems ($nc\bar{n}\bar{c}$ and $nb%
\bar{n}\bar{b}$)}

\begin{figure}[th]
\centering
\includegraphics[width=0.5\textwidth]{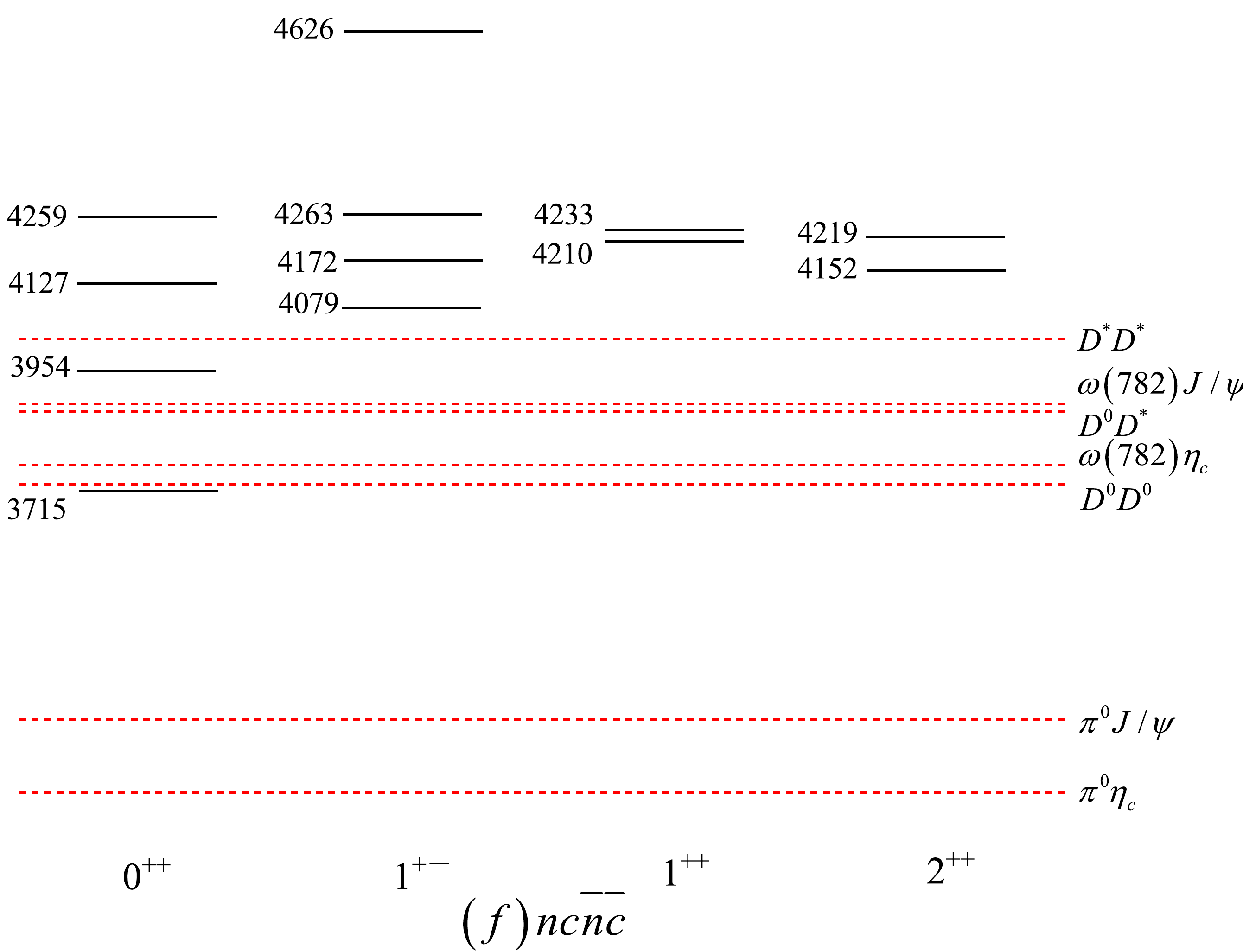}
\caption{Computed masses(MeV the solid lines) of the hidden-bottom tetraquarks $%
nc\bar{n}\bar{c}$ and the two meson thresholds (MeV the dotted lines). }
\label{nc}
\end{figure}
\begin{figure}[th]
\centering
\includegraphics[width=0.5\textwidth]{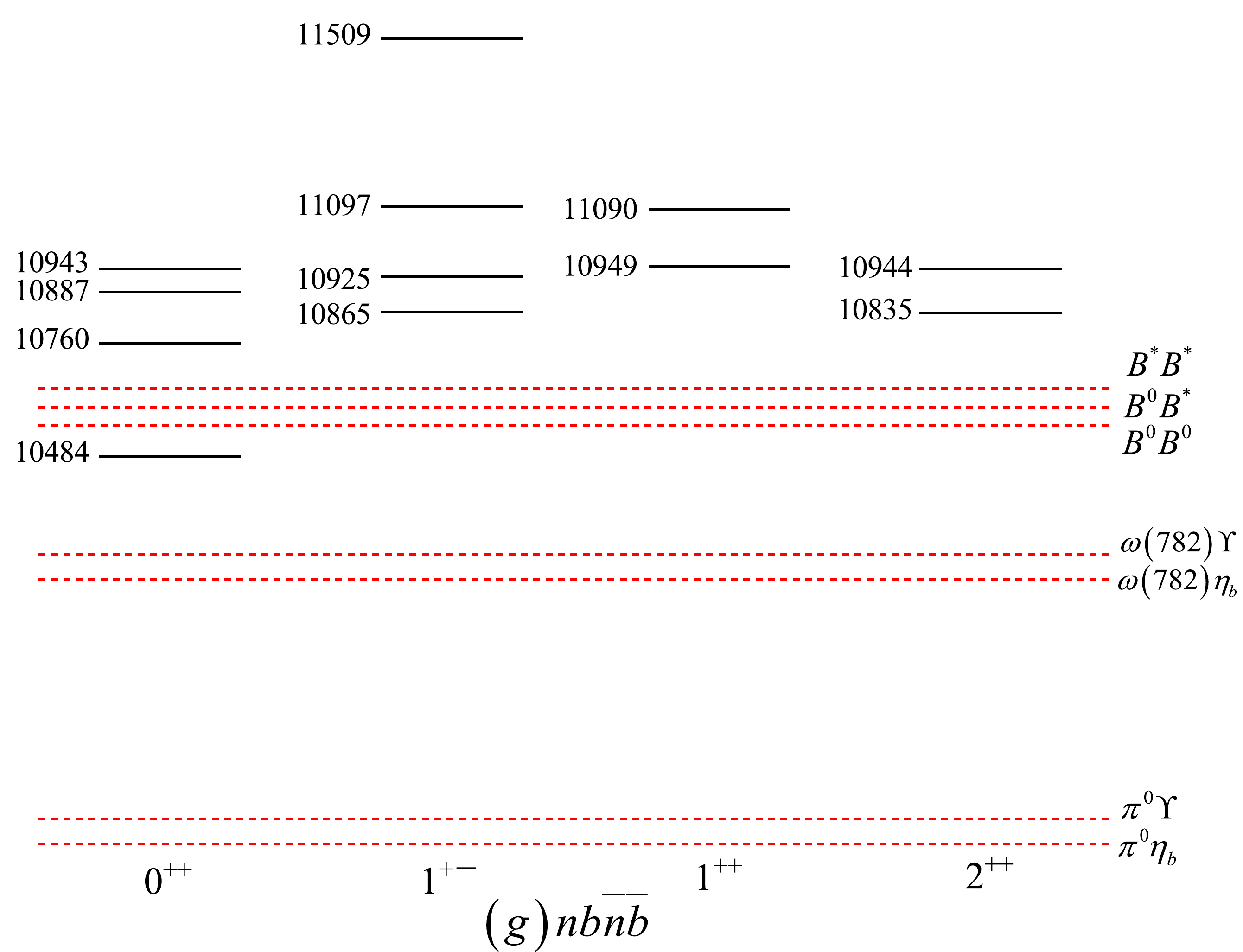}
\caption{Computed masses(MeV the solid lines) of the hidden-bottom tetraquark $%
nb\bar{n}\bar{b}$ and the two meson thresholds (MeV the dotted lines). }
\label{nb}
\end{figure}

For hidden charmed systems of the tetraquarks $nc\bar{n}\bar{c}$, we show
the computed results for $R_{0}$, the mixing weights, the masses $M(T)$ and
thresholds in Table \ref{tab:ncncnbnbmass}, with the later two plotted in
Fig \ref{nc}. There are four states for each of $J^{PC}=0^{++}$ and $%
J^{PC}=1^{+-}$, and two states for each of $J^{PC}=1^{++}$ and $%
J^{PC}=2^{++} $. For the $0^{++}$ states, two higher states ($4259$ MeV, $%
4127\,$MeV) are all above the thresholds of $D^{0}D^{0}$, $D^{\ast }D^{\ast
} $, $\omega \left( 782\right) J/\psi $ and $\pi ^{0}\eta _{c}$ (about $%
397-529\,$MeV, $110-242\,$MeV, $248-380\,$MeV and $1008-1140\,$MeV). The
lower state with mass $3954\,$MeV, which is above the thresholds of $%
D^{0}D^{0}$, $\omega \left( 782\right) J/\psi $, $\pi ^{0}\eta _{c}$ and
below the threshold of $D^{\ast }D^{\ast }$, can strongly decay to the three
former final states. The lowest state, which is below the thresholds of $%
D^{\ast }D^{\ast }$, $\omega \left( 782\right) J/\psi $,$D^{0}D^{0}$ and
above the thresholds of $\pi ^{0}\eta _{c}$, can decay to two final states
of $\pi ^{0}\eta _{c}$ . Further, all states with $J^{PC}=1^{+-}$, $1^{++}$
and $2^{++}$ are above the thresholds of $D^{0}D^{\ast }$, $D^{\ast }D^{\ast
}$, $\pi ^{0}J/\psi $, $\omega \left( 782\right) J/\psi $, $\omega \left(
782\right) \eta _{c}$, can decay to the laters with same quantum numbers.
For instance, the $1^{++}$ states can decay to $\omega \left( 782\right)
J/\psi $ and $D^{0}D^{\ast }$, the $2^{++} $ states can decay to $D^{\ast
}D^{\ast }$, $\omega \left( 782\right)$$J/\psi$.

For hidden-bottom systems of tetraquarks $nb\bar{n}\bar{b}$, we show the
computed results for $R_{0}$, the mixing weights, the masses $M(T)$ and
thresholds in Table \ref{tab:ncncnbnbmass}, with the later two plotted in
Fig \ref{nb}. We find from Fig \ref{nb} that all states of $nb\bar{n}\bar{b}$
systems are above the thresholds of their final states of two mesons, except
for the lowest state ($10484\,$MeV), which is below the thresholds of $%
B^{0}B^{0}$ and $B^{\ast }B^{\ast }$ only and it can decay into $\omega
\left( 782\right) \Upsilon $, $\pi ^{0}\eta _{b}$. The possible decays are,
for instance, the $nb\bar{n}\bar{b}(0^{++})$ to $B^{0}B^{0}$, $B^{\ast
}B^{\ast }$, $\omega \left( 782\right) \Upsilon $ and $\pi ^{0}\eta _{b}$,
the $nb\bar{n}\bar{b}(1^{+-})$ to $B^{0}B^{\ast }$, $B^{\ast }B^{\ast }$, $%
\pi ^{0}\Upsilon $ and $\omega \left( 782\right) \eta _{b}$, the $nb\bar{n}%
\bar{b}(1^{++})$ to $B^{0}B^{\ast }$, $\omega \left( 782\right) \Upsilon $,
the $nb\bar{n}\bar{b}(2^{++})$ to $B^{\ast }B^{\ast }$, $\omega \left(
782\right) \Upsilon $. \renewcommand{\tabcolsep}{0.35cm} \renewcommand{%
\arraystretch}{2.05}
\begin{table*}[tbh]
\caption{Computed results for the hidden heavy-flavor tetraquark $nc\bar{n}%
\bar{c}$ and $nb\bar{n}\bar{b}$, with respective thresholds of two mesons
shown also. }
\label{tab:ncncnbnbmass}%
\begin{tabular}{ccccccc}
\hline\hline
\textrm{State} & $J^{PC}$ & \textrm{Eigenvector} & $R_{0}$(GeV$^{-1}$) & $%
M(T)$(MeV) & \textrm{Threshold }(MeV) &  \\ \hline
$nc\bar{n}\bar{c}$ & $0^{++}$ & $(-0.24,-0.45,0.83,0.20)$ & $4.90$ & $3715$
& $D^{0}D^{0}=3730$;$D^{\ast }D^{\ast }=4017$ &  \\
&  & $(-0.71,0.26,-0.22,0.62)$ & $5.01$ & $3954$ & $\omega \left( 782\right)
J/\psi =3879$ &  \\
&  & $(0.17,0.85,0.50,0.00)$ & $5.17$ & $4127$ & $\pi ^{0}\eta _{c}=3119$ &
\\
&  & $(0.65,-0.11,-0.06,0.75)$ & $5.38$ & $4259$ &  &  \\
& $1^{+-}$ & $(-0.31,0.65,-0.06,0.69)$ & $5.15$ & $4079$ & $D^{0}D^{\ast
}=3874$;$D^{\ast }D^{\ast }=4017$ &  \\
&  & $(0.87,0.33,0.34,0.11)$ & $5.26$ & $4172$ & $\pi ^{0}J/\psi =3264$;$%
\omega \left( 782\right) \eta _{c}=3767$ &  \\
&  & $(-0.15,-0.50,0.70,0.48)$ & $5.33$ & $4263$ &  &  \\
&  & $(0.35,-0.46,-0.61,0.54)$ & $5.32$ & $4626$ &  &  \\
& $1^{++}$ & $(-0.58,0.82)$ & $5.18$ & $4210$ & $D^{0}D^{\ast }=3874$;$%
\omega \left( 782\right) J/\psi =3879$ &  \\
&  & $(0.82,0.58)$ & $5.36$ & $4233$ &  &  \\
& $2^{++}$ & $(0.46,0.89)$ & $5.34$ & $4152$ & $D^{\ast }D^{\ast }=4017$;$%
\omega \left( 782\right) J/\psi =3879$ &  \\
&  & $(-0.89,0.46)$ & $5.31$ & $4219$ &  &  \\
$nb\bar{n}\bar{b}$ & $0^{++}$ & $(-0.37,-0.31,0.80,0.36)$ & $4.65$ & $10484$
& $B^{0}B^{0}=10560$;$B^{\ast }B^{\ast }=10650$ &  \\
&  & $(-0.51,0.37,-0.40,0.67)$ & $4.70$ & $10760$ & $\omega \left(
782\right) \Upsilon =10242$ &  \\
&  & $(0.42,0.78,0.43,0.14)$ & $4.82$ & $10887$ & $\pi ^{0}\eta _{b}=9533$ &
\\
&  & $(0.65,-0.40,-0.14,0.63)$ & $4.90$ & $10943$ &  &  \\
& $1^{+-}$ & $(-0.43,0.67,-0.25,0.54)$ & $4.79$ & $10865$ & $B^{0}B^{\ast
}=10605$;$B^{\ast }B^{\ast }=10650$ &  \\
&  & $(0.73,0.41,0.47,0.29)$ & $4.84$ & $10925$ & $\pi ^{0}\Upsilon =9594$;$%
\omega \left( 782\right) \eta _{b}=10182$ &  \\
&  & $(-0.36,-0.45,0.61,0.55)$ & $4.81$ & $11097$ &  &  \\
&  & $(0.39,-0.42,-0.59,0.57)$ & $4.95$ & $11509$ &  &  \\
& $1^{++}$ & $(0.82,0.58)$ & $4.89$ & $10949$ & $B^{0}B^{\ast }=10605$;$%
\omega \left( 782\right) \Upsilon =10242$ &  \\
&  & $(-0.58,0.82)$ & $4.78$ & $11090$ &  &  \\
& $2^{++}$ & $(0.45,0.89)$ & $4.88$ & $10835$ & $B^{\ast }B^{\ast }=10650$;$%
\omega \left( 782\right) \Upsilon =10242$ &  \\
&  & $(-0.89,0.45)$ & $4.88$ & $10944$ &  &  \\ \hline\hline
\end{tabular}%
\end{table*}

\section{Summary}

Stimulated by observations of the $X(6900)$ by LHCb and the recent
observations of the $X(6600)$ by CMS and ATLAS experiments of the LHC, we
have systematically investigated the ground-state masses of hidden
heavy-flavor tetraquarks with two and four hidden heavy-flavor within a
unified framework of MIT bag model which incorporates chromomagnetic
interactions and enhanced binding energy. Based on color-spin wavefunctions
constructed for the hidden heavy-flavor tetraquarks, we solve the MIT bag
model and diagonalize the chromomagnetic interaction (CMI) to predict masses
of the color-spin multiplets of hidden heavy-flavor tetraquarks in their
ground states with spin-parity quantum numbers $J^{PC}=0^{++}$, $1^{++}$, $%
2^{++}$, and $1^{+-}$. We find that the fully charmed tetraquark $cc\bar{c}%
\bar{c}$ with $J^{PC}=0^{++}$ has mass about $6572$ MeV and is very likely
to be the $X(6600)$ reported by CMS and ATLAS experiments of the LHC, with
the measured mass $6552\pm 10\pm 12$ MeV. We further computed masses of the
tetraquark systems $bb\bar{b}\bar{b}$, $cb\bar{c}\bar{b}$, $sc\bar{s}\bar{c}$%
, $sb\bar{s}\bar{b}$, $nc\bar{n}\bar{c}$ and $nb\bar{n}\bar{b}$ in their
color-spin multiplets and suggested that the particle $Z_{c}(4200)$ reported
by \cite{Belle:2014nuw} is likely to be the hidden-charm tetraquark made of $%
nc\bar{n}\bar{c}$ with $J^{PC}=1^{+-}$.

Compared to two-meson thresholds determined via the final states in details,
the most-likely strong decay channels are noted. Our mass computation shows
that all of these hidden heavy-flavor tetraquarks are above the thresholds of the
lowest two-mesons final states and unstable against strong decay to these
final states. For the doubly heavy systems of the tetraquarks $sb\bar{s}\bar{%
b}$, $sc\bar{s}\bar{c}$, $nb\bar{n}\bar{b}$ and $nc\bar{n}\bar{c}$, there
are a few states below thresholds except for their lowest final states,
indicating that they may have longer lifetime compared to the fully heavy
tetraquarks. We also find some near-threshold states for which coupled
channel effects are possible. We hope that upcoming LHCb experiments with
increased data can test the prediction in this work.

\textbf{Acknowledgments}

D. J. is supported by the National Natural Science Foundation of China under
the no. 12165017.

\section*{\textbf{Appendix A}}

Based on the color $SU(3)_{c}$ symmetry, one can obtain two components of
color singlets $6_{c}\otimes \bar{6}_{c}$ and $\bar{3}_{c}\otimes 3_{c}$ for
the hidden-flavor tetraquarks,
\begin{eqnarray}
\phi _{1}^{T} &=&\frac{1}{\sqrt{6}}\left( rr\bar{r}\bar{r}+gg\bar{g}\bar{g}%
+bb\bar{b}\bar{b}\right)  \notag \\
&&+\frac{1}{2\sqrt{6}}\left( rb\bar{b}\bar{r}+br\bar{b}\bar{r}+gr\bar{g}\bar{%
r}+rg\bar{g}\bar{r}+gb\bar{b}\bar{g}+bg\bar{b}\bar{g}\right.  \notag \\
&&+gr\bar{r}\bar{g}+rg\bar{r}\bar{g}+gb\bar{g}\bar{b}\left. +bg\bar{g}\bar{b}%
+rb\bar{r}\bar{b}+br\bar{r}\bar{b}\right) ,  \label{ph1}
\end{eqnarray}%
\begin{eqnarray}
\phi _{2}^{T} &=&\frac{1}{2\sqrt{3}}\left( rb\bar{b}\bar{r}-br\bar{b}\bar{r}%
-gr\bar{g}\bar{r}+rg\bar{g}\bar{r}+gb\bar{b}\bar{g}-bg\bar{b}\bar{g}\right.
\notag \\
&&\left. +gr\bar{r}\bar{g}-rg\bar{r}\bar{g}-gb\bar{g}\bar{b}+bg\bar{g}\bar{b}%
-rb\bar{r}\bar{b}+br\bar{r}\bar{b}\right) ,  \label{ph2}
\end{eqnarray}%
which corresponds to two color configurations in Eq. (\ref{colorT}).

For six states $\chi _{1\sim 6}^{T}$ (\ref{spinT}) of heavy tetraquarks, one
can construct their spin wave functions via writing the $CG$ coefficients
explicitly:%
\begin{align}
\chi_{1}^{T}&=\uparrow\uparrow\uparrow\uparrow, \nonumber \\
\chi_{2}^{T}&=\frac{1}{2}\left(\uparrow\uparrow\uparrow\downarrow+\uparrow\uparrow\downarrow\uparrow-%
\uparrow\downarrow\uparrow\uparrow-\downarrow\uparrow\uparrow\uparrow%
\right), \nonumber \\ \chi_{3}^{T}&=\frac{1}{\sqrt{3}}
\left(\uparrow\uparrow\downarrow\downarrow+\downarrow\downarrow\uparrow%
\uparrow\right), \nonumber \\ &-\frac{1}{2\sqrt{3}}
\left(\uparrow\downarrow\uparrow\downarrow+\uparrow\downarrow\downarrow%
\uparrow+\downarrow\uparrow\uparrow\downarrow+\downarrow\uparrow\downarrow%
\uparrow\right), \nonumber \\ \chi_{4}^{T}&=\frac{1}{\sqrt{2}}
\left(\uparrow\uparrow\uparrow\downarrow-\uparrow\uparrow\downarrow\uparrow%
\right), \nonumber \\ \chi_{5}^{T}&=\frac{1}{\sqrt{2}}
\left(\uparrow\downarrow\uparrow\uparrow-\downarrow\uparrow\uparrow\uparrow%
\right), \nonumber \\ \chi_{6}^{T}&=\frac{1}{2}
\left(\uparrow\downarrow\uparrow\downarrow-\uparrow\downarrow\downarrow%
\uparrow-\downarrow\uparrow\uparrow\downarrow+\downarrow\uparrow\downarrow%
\uparrow\right), \label{pspinT}
\end{align}
in which notations $\uparrow $ and $\downarrow $ represent the third
component of the quark's spin. Alternatively, one can also use the $CG$
coefficients given in Ref. \cite{Zhang:2021yul} to examine Eq. (\ref{pspinT}%
) for the different spin states. Note that the results of spin factors in
Ref. \cite{Zhang:2021yul} are shown as matrix form in the spanned space of
the states $\chi _{1\sim 6}^{T}$ as the spin multiplets (\ref{spinT})
indicated. Combining with two color configurations $\phi _{1\sim 2}^{T}$ in
Eq. (\ref{colorT}) and six spin configurations $\chi _{1\sim 6}^{T}$ in Eq. (%
\ref{spinT}), one can then construct their color-spin wavefunctions (\ref%
{colorspinT}). The allowed states of the hidden-flavor tetraquarks to be mix
due to chromomagnetic interaction are listed in Table 1.

\end{document}